\documentclass[fleqn,usenatbib]{mnras}

\usepackage{gensymb}
\usepackage[T1]{fontenc}
\usepackage{ae,aecompl}
\usepackage{siunitx}

\usepackage{graphicx}	
\usepackage{amsmath}	
\usepackage{amssymb}	
\usepackage{float}
\usepackage{lineno}
\usepackage{tabularx}
\usepackage{rotating}

\title[TON\,0599 MAGIC detection]{First detection of VHE gamma-ray signal from the FSRQ TON\,0599}

\author[K.~Abe~et.~al.]{\parbox{\textwidth}{\Large{
MAGIC Collaboration: K.~Abe$^{1}$,
S.~Abe$^{2}$,
J.~Abhir$^{3}$,
A.~Abhishek$^{4}$,
V.~A.~Acciari$^{5}$,
A.~Aguasca-Cabot$^{6}$,
I.~Agudo$^{7}$,
T.~Aniello$^{8}$,
S.~Ansoldi$^{9,42}$,
L.~A.~Antonelli$^{8}$
% ...
G.~A.~Borman$^{50}$, 
W.~Boschin$^{15,51}$, 
V.~Bozhilov$^{52,53}$, 
M.~I.~Carnerero$^{54}$, 
D.~Carosati$^{51,55}$, 
C.~Casadio$^{56}$, 
W.~Chamani$^{57}$, 
W.~P.~Chen$^{58}$, 
G.~Damljanovic$^{59}$, 
S.~A.~Ehgamberdiev$^{60,61}$, 
K.~E.~Ergashev$^{60}$, 
M.~Giroletti$^{62}$, 
T.~S.~Grishina$^{63}$, 
M.~A.~Gurwell$^{64}$, 
M.~K.~Hallum$^{65}$, 
E.~J\"arvel\"a$^{66}$, 
H.~E.~Jermak$^{67}$, 
S.~G.~Jorstad$^{63,65}$, 
M.~D.~Jovanovic$^{59}$, 
E.~N.~Kopatskaya$^{63}$, 
K.~Kuratov$^{68,69}$\thanks{Recently deceased.}, 
O.~M.~Kurtanidze$^{70,71,72,73}$, 
S.~O.~Kurtanidze$^{70,72}$, 
A.~L\"ahteenm\"aki$^{74,75}$, 
E.~G.~Larionova$^{63}$, 
L.~V.~Larionova$^{63}$, 
V.~M.~Larionov$^{63,76}$, 
H.~C.~Lin$^{58}$, 
N.~Marchili$^{62}$, 
A.~P.~Marscher$^{65}$, 
C.~McCall$^{67}$, 
M.~Minev$^{77}$, 
D.~O.~Mirzaqulov$^{60}$, 
I.~Myserlis$^{71,78}$, 
M.~G.~Nikolashvili$^{70,72}$, 
T.~Pursimo$^{79,80}$, 
C.~M.~Raiteri$^{54}$, 
I.~Reva$^{81}$, 
S.~Righini$^{62}$, 
A.~C.~Sadun$^{82}$, 
S.~S.~Savchenko$^{63,76}$, 
E.~Semkov$^{77}$, 
P.~S.~Smith$^{83}$, 
I.~A.~Steele$^{67}$, 
M.~Tornikoski$^{74}$, 
Yu.~V.~Troitskaya$^{63}$, 
I.~S.~Troitskiy$^{63}$, 
A.~A.~Vasilyev$^{63}$, 
M.~Villata$^{54}$, 
O.~Vince$^{59}$, 
Z.~R.~Weaver$^{65}$ \\
\emph{\normalsize Affiliations are listed at the end of the paper}
}}}

\date{Accepted XXX. Received YYY; in original form ZZZ}

\pubyear{2026}

\begin{document}
\label{firstpage}
\pagerange{\pageref{firstpage}--\pageref{lastpage}}
\maketitle

\newpage
\begin{abstract}
TON\,0599 (z=0.7247) belongs to the few flat spectrum radio quasars (FSRQs) detected in the very high energy (VHE, $E > 100$\,GeV) gamma-ray band. Its redshift makes it currently one of the farthest VHE gamma-ray sources.
It was detected for the first time with the MAGIC telescopes on 2017 December 15, and observed until December 29. The flux reached a maximum of about 50 per cent of the Crab Nebula flux above 80 GeV on the second night of observation, after which we witnessed a gradual decrease of the flux. 
The VHE gamma-ray spectrum connects smoothly to the one in the high energy ($E > 100$\,MeV) band obtained from simultaneous observations with {\textit Fermi}-LAT. It features a cut-off at energies around 50\,GeV, indicating the location of the gamma-ray emission zone beyond the broad line region. In addition, we were able to follow the spectral evolution during the fading phase of the flare. 
Multiwavelength analysis based on observations in optical, 
near-infrared, and radio bands acquired by the Whole Earth Blazar Telescope 
 (WEBT) Collaboration 
from November to March, as well as observations in X-ray and 
optical--UV bands with instruments on board the \textit{Swift} satellite, shows strong correlation between different bands.
We model the broadband emission with a simple one-zone leptonic model, where the high-energy peak is predominantly produced by external Compton (EC) scattering of photons from the dusty torus.

\end{abstract}




\section{Introduction}
Blazars are a subclass of active galactic nuclei (AGN) which eject material through collimated jets of ultrarelativistic matter \citep{ur95}. 
They are known for significant and fast flux variability, with time scales as short as order of minutes. Blazars are subdivided into BL Lacertae objects (BL Lacs) and flat spectrum radio quasars (FSRQs). Although BL Lacs generally have featureless optical spectra, FSRQs 
usually exhibit prominent emission lines and strong continuum emission in the optical-UV band (the so-called blue bump) associated with the emission
from the accretion disc. Although FSRQs generally have higher luminosities than BL Lacs, their inverse Compton (IC) bumps peak at lower energies, resulting in softer very high energy (VHE, $E>100$\,GeV) spectra. In addition, they are mostly located at higher redshifts \citep{ackermann2011}, with their VHE spectra further softened by absorption on extragalactic background light (EBL). For these reasons, FSRQs are more difficult to detect in the VHE band, TON\,0599 being one of the handful detected so far.

TON\,0599 (other names: 4C\,+29.45, B2\,1156+29, TeV\,J1159+292, RGB\,J1159+292, RX\,J1159.5+2914) is located at Right Ascension (J2000.0) 11h59m31.8s and Declination +29d14m44s. At the redshift of $z = 0.7247$ \citep{he10} it is currently the sixth farthest source of VHE gamma rays, and fifth farthest AGN. 
TON\,0599 is a highly variable source in all frequencies~\citep{1983ApJ...274...62W, 1992ApJ...398..454W, 2007A&A...469..899H, 2008A&A...489L..33S, 2013A&A...555A.134L, 2014MNRAS.443...58W}.
\citet{Stockman1978} and \citet{fa06} classified it as an optically violently variable (OVV) quasar based on polarization and photometric measurements, respectively. 
In addition, \citet{fa06} found possible periodicity in the optical light curve with time scales of $1.58$ and $3.55$ years. 
The flux variability in near-infrared was followed in 2007--2013 at the Campo Imperatore and Teide observatories, monitored by the GLAST-AGILE Support Program (GASP) of the Whole Earth Blazar Telescope (WEBT) Collaboration.
The fractional variability (see Section~\ref{subsec:Fvar}) in \textit{J} band in that period was 0.73 \citep{ra14}.
Also, intranight variability at the level of 22.3 per cent of the flux was observed with Sampurnanand Telescope, possibly caused by turbulence in the jet~\citep{2013MNRAS.435.1300G}.

The first detection of TON\,0599 in gamma rays was reported in the second EGRET catalogue \citep{thompson1995}. It was also listed in the First {\textit Fermi} Large Area Telescope (LAT) source catalogue \citep{abdo2010}. 
TON\,0599 underwent a particulary active flaring state in late 2017. 
Multiwavelength (MWL) studies analysing that period were published in \citet{patel2018, prince2019, 2022ApJ...926..180H, 10.1093/mnras/stz3490}.

On 2017 December 15, TON\,0599 was detected for the first time in the VHE band with the Major Atmospheric Gamma Imaging Atmospheric (MAGIC) telescopes \citep{Mirzoyan2017}. The details of the detection are reported in this work. In Section~\ref{sec:analysis}, we describe the observations used in this study and lay out the results. The MWL picture is presented in Section~\ref{sec:MWL}, and the results of the broadband modelling in Section~\ref{subsec:modelling}. We summarize our results in Section~\ref{sec:conclusions}.

\section{Data analysis and results}
\label{sec:analysis}

\subsection{Gamma rays}
\label{subsec:gamma-rays}	

\subsubsection{MAGIC observation and analysis}
\label{subsec:magic} 
MAGIC is a system of two $17\,$m mirror dish Imaging Atmospheric Cherenkov Telescopes (IACTs) located in El Roque de los Muchachos observatory in Canary Island of La Palma ($28.7^\circ$\,N, $17.9^\circ$\,W), at the altitude of 2200 m a.s.l. \citep{al16a}. 
MAGIC observations of TON\,0599 started on 2017 December 15 (MJD 58102), following a hardening of the HE gamma-ray spectrum observed with \textit{Fermi}-LAT. The observations lasted until 2017 December 29 (MJD 58116). During this time, 19.2 hours of data were collected. All observations were performed in the so-called {\it wobble} mode \citep{fomin94,al16b}, under dark sky conditions \citep{al16a}. 
Based on measurements with a pyrometer and an AllSky camera \citep{will2017} and subjective estimation by the observers\footnote{The MAGIC LIDAR system used for monitoring the atmospheric transmission \citep{fruck2015} was not functioning properly during the campaign.}, the atmospheric transmission was estimated to be above 85 per cent with respect to clean atmosphere for all observations.
The data were reduced and analysed using MAGIC Analysis and Reconstruction Software (MARS, \citealt{za13}). 
For points whose relative uncertainties are greater than 0.5, the upper limits on the flux have been computed following \cite{ro05} at the confidence level of 95 per cent.

\begin{figure}
    \centering
	\includegraphics[width=\columnwidth]{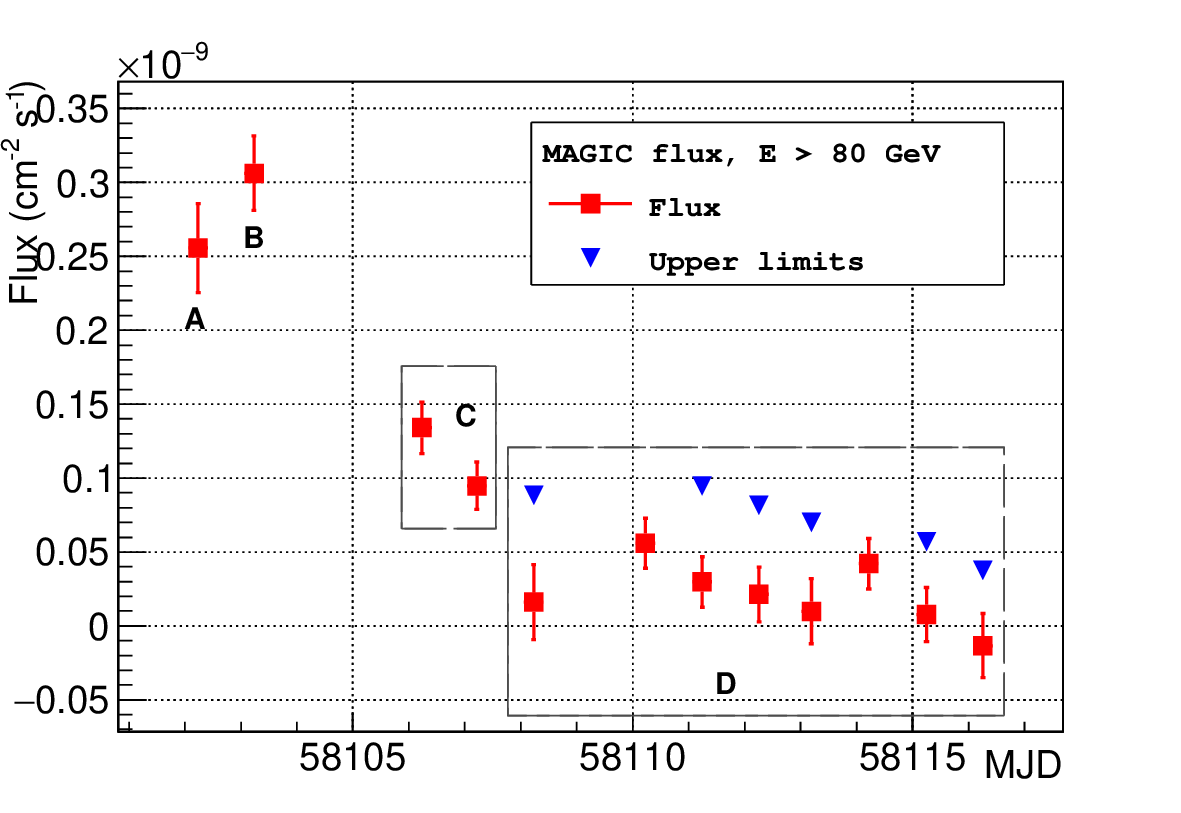}
    \caption{Integrated flux measured by MAGIC above 80\,GeV. Blue triangles represent upper limits for points whose relative uncertainties are greater than 0.5. Letters ``A", ``B", ``C", and ``D" mark the four periods, as described in the main text.}
    \label{fig:LC_MAGIC}
\end{figure}

On the first night, after observing for 0.96 hours with the MAGIC telescopes, the signal was detected with a statistical significance of 10.4\,$\sigma$ (using Eq. 17 of \citealt{LiMa83}). 
The integral flux above 80\,GeV reached $(2.6 \pm 0.3)\times 10^{-10}$\,cm$^{-2}$\,s$^{-1}$, corresponding to 42 per cent of flux from the Crab Nebula (0.42 C.U.) in the same energy range, as reported in \citet{al15a}. 
The second night showed a marginal increase in flux (although within uncertainties) to $(3.1 \pm 0.3)\times 10^{-10}$\,cm$^{-2}$\,s$^{-1}$, which corresponds to 0.51 C.U. above 80\,GeV. 
After the second night, the measured flux decreased until the end of the MAGIC observation campaign. The light curve above 80\,GeV is shown in Fig.~\ref{fig:LC_MAGIC}, while in Fig.~\ref{fig:LC_MWL} it is put in the MWL context.

\subsubsection{\textit{Fermi}-LAT observation and analysis}
\label{subsec:Fermi}	
Observations of the TON\,0599 VHE flare were triggered by applying the \textit{iSRS} clustering scheme \citep{pacciani2018} on incoming \textit{Fermi}-LAT data \citep{atwood2009} for gamma-ray energies above 10 GeV. 
The triggering procedure runs on the entire sample of blazars within the third \textit{Fermi}-LAT catalog~\citep{Acero_2015}. For each source, gamma rays are collected if contained within a circular region centred on the source and with a radius $R^{95}$ corresponding to the 95 per cent containment ($R^{95}$ depends on the energy and on the reconstruction category of the incoming gamma ray). 
The \textit{iSRS} procedure was adopted to trigger observations of several other flares from the FSRQs and BL Lacs~\citep{pacciani2014,ahnen2015,ansoldi2018}.\\ 
Spectra and binned light curves were produced using the standard analysis methods for PASS8 data (Science-Tools \texttt{v10r0p5}, provided by the \textit{Fermi}-LAT collaboration), and the Pass8R2 V6 Source instrument response function, selecting events
of class P8 SOURCE.\\
We extracted gamma-ray events within 20$\degree$ from the source position and selected events of class \textit{P8 Source}. Events from the Earth limb were rejected adopting a zenith angle cut of 90$\degree$. 
The source flux in the energy and temporal bins was extracted using the unbinned likelihood analysis \citep{cash1979,abdo2009}. 
Likelihood analysis made use of \texttt{p8r2\_source\_v6\_\_gll\_iem\_v06} galactic diffuse background model and of \texttt{p8r2\_source\_v6\_\_iso\_source\_v06} isotropic diffuse background model.\\
The \textit{Fermi}-LAT light curve, obtained in the energy range 300 MeV -- 500 GeV, is reported with 24 hour time bins. Upper limits are shown for time bins with
source test statistics (TS) $< 9$.  
We evaluated that the photon indices obtained by fitting the data with a power-law could be unreliable in the case of low source statistics, resulting in a softer photon index than expected. Simulating sources with the power-law spectra of photon index 2.2, the distribution of reconstructed photon index has a tail toward soft-photon indices.
The reconstructed photon index is $>3$ in $\sim 19$ per cent (8 per cent) of the cases for 6 (12) source counts (npred). 
The reconstructed source photon index for the 24 hour time bins and source TS > 9 is reported in Fig.~\ref{fig:LC_MWL}. 
The time bins with npred\,$\ge$\,12 are highlighted in red. 
We checked the effect of applying an incorrect weight for the isotropic and galactic background. Assuming the total background contribution to be overestimated/underestimated by 20 per cent, the effect is negligible both in spectral shape and the flux estimate for periods A and B. For period C (period D), the background overestimation/underestimation would result in hardening/softening of the photon index of 0.014 (0.018), and to an underestimation/overestimation of the flux by 2.2 per cent (4.2 per cent). 

The fastest variability in the HE flux was observed between MJD 58052 and 58061. We fitted the flux with a Gaussian function. The mean of the Gaussian is located at MJD $58057.6\pm 0.2$ with a standard deviation of $2.7 \pm 0.2$ days. 

\subsubsection{Gamma-ray results}
\label{subsec:gamma-ray_results}	
\begin{figure*}
    \centering
	\includegraphics[width=\columnwidth]{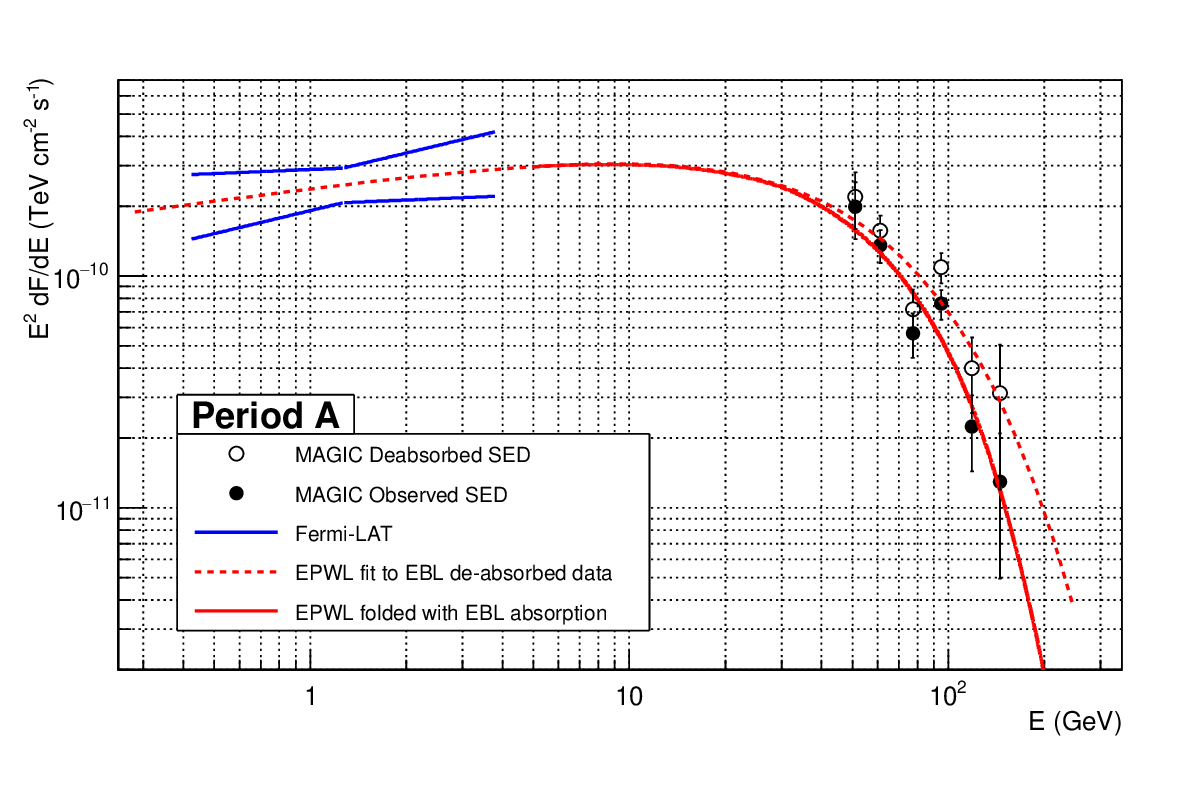}
    \includegraphics[width=\columnwidth]{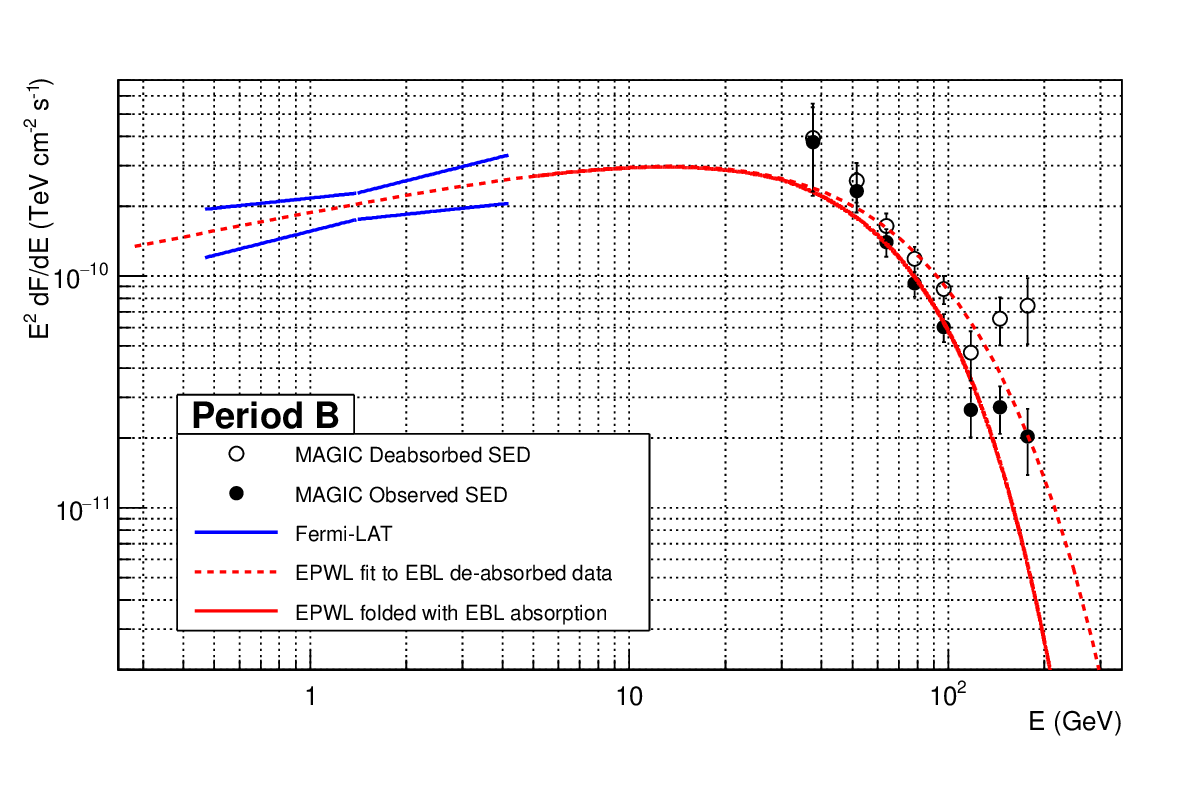}
    \includegraphics[width=\columnwidth]{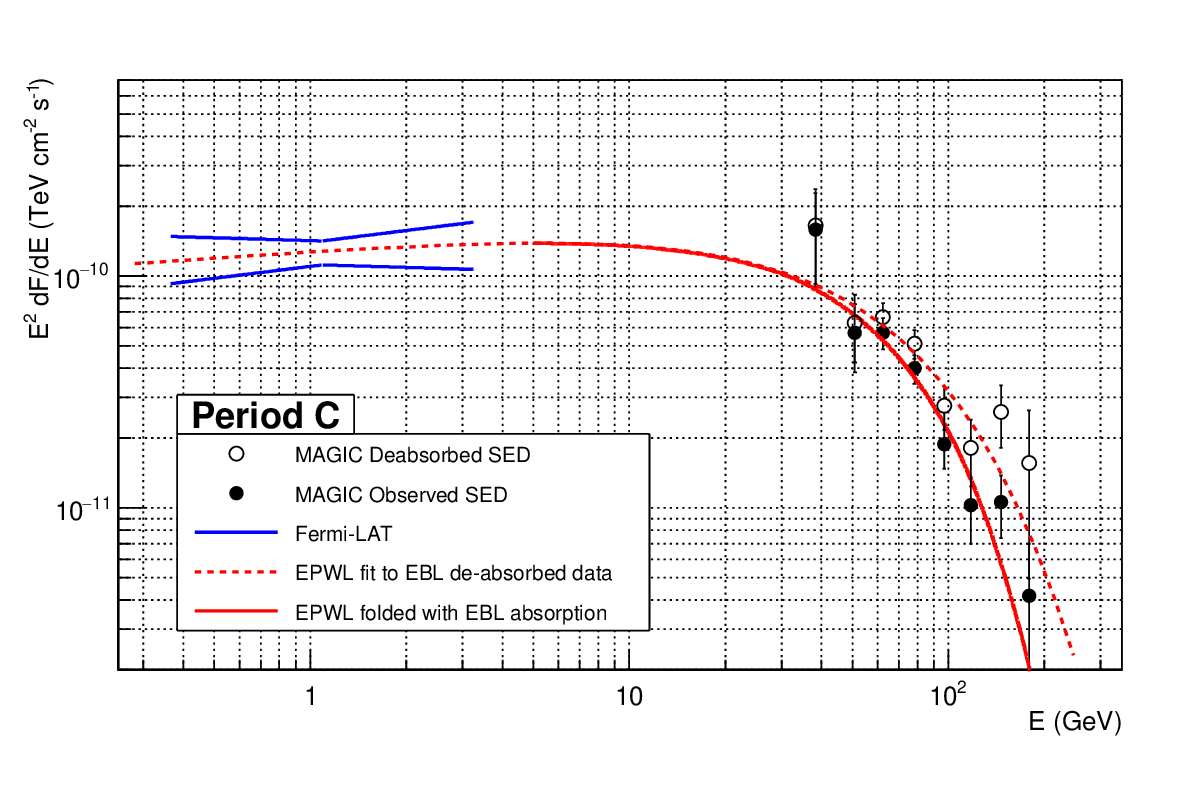}
    \includegraphics[width=\columnwidth]{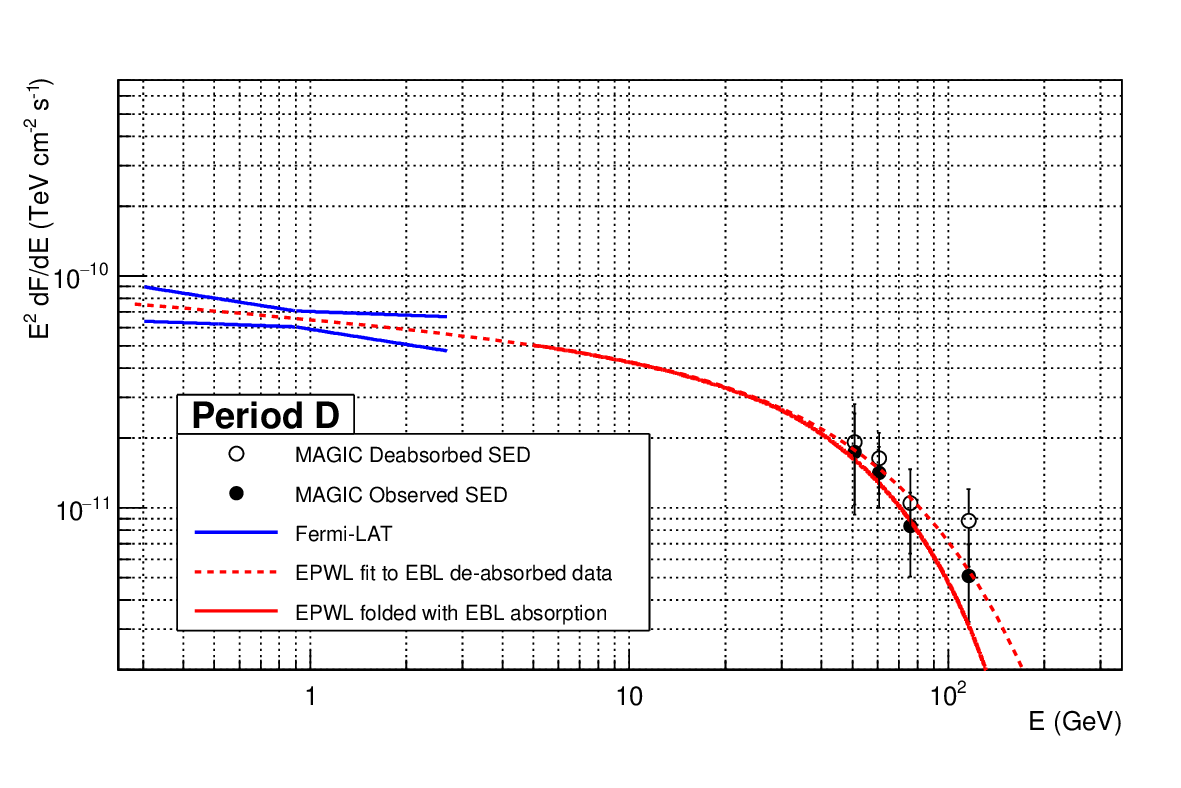}
    \caption{Spectral energy distributions in the gamma-ray band for periods A through D. Full black circles are MAGIC measurements. Only points with excess significance greater than 2 are displayed. Black empty circles represent intrinsic (EBL de-absorbed) SED values. Blue butterflies represent \textit{Fermi}-LAT SEDs, with parameters given in Table~\ref{tab:LAT_butterflies}. Source intrinsic SEDs are represented with dashed red lines. The results of the fit are given in Table~\ref{tab:MAGIC-LAT_FitResults}. Solid red lines represent the intrinsic spectra attenuated by the EBL absorption.} 
    \label{fig:MAGIC-LAT_SED}
\end{figure*}

The data were divided in four sub-sets according to the VHE gamma-ray flux level: the first two nights of observation are considered to be the `high' flux level nights, and are referred to as Period A and B, respectively; Period C contains `intermediate' flux level nights December 19 and 20 (3rd and 4th point in the MAGIC light curve); Period D contains all other nights, i.e. December 21 to 29, and represents the period of `low' flux level.
We calculated the spectra for each of the four periods separately, combining MAGIC data with simultaneous data from \textit{Fermi}-LAT. The SED plots for each period are shown in Fig.~\ref{fig:MAGIC-LAT_SED}. 
The MAGIC measured spectral points (represented with full black circles) were de-absorbed to account for absorption of gamma rays on EBL, using the model from \citet{dom11}. These points are represented with empty black circles. 
The \textit{Fermi}-LAT data were integrated over 24 hours for each day of a given period; e.g., Period B was integrated from December 15 at 16:48 hours until December 16 at 16:48 hours, while Period D was integrated from December 20 at 16:48 hours till December 29 at 16:48 hours. 
\textit{Fermi}-LAT photon index shows a hardening trend between MJD 58100 and 58103 (see Fig.~\ref{fig:LC_MWL}). Therefore, we chose a shorter integration time for Period A. Data were integrated over 12 hours (December 14 at 22:48 hours until December 15 at 10:48 hours) in order to keep the uncertainties relatively small. In all four periods, the \textit{Fermi}-LAT spectra were fitted with a power-law.  
The fitted \textit{Fermi}-LAT spectra are represented with blue butterflies in Fig.~\ref{fig:MAGIC-LAT_SED}, with the parameters given in Table~\ref{tab:LAT_butterflies}. 
\begin{table*}
	\centering
    \caption{Spectral parameters of the \textit{Fermi}-LAT butterflies shown in Fig.~\ref{fig:MAGIC-LAT_SED}. Data given in columns is as follows: Sub-set ID; integrated flux between 300\,MeV and 500\,GeV; decorrelation energy; photon index, SED at the decorrelation energy.}
	\label{tab:LAT_butterflies}
    \begin{tabular}{ccccc}
		\hline
		Period & F [cm$^{-2}$ s$^{-1}$] & $E_\mathrm{dec}$ [GeV] & ph. index & $\nu f_{\nu} (E_\mathrm{dec})$ \\
         & ($0.3 < E < 500$\,GeV) & & $\alpha$ & [erg cm$^{-2}$ s$^{-1}$] \\
		\hline
		A & $(7.8 \pm 1.3)\times 10^{-7}$ & 1.27 & $-1.81 \pm 0.14$ & $(4.0 \pm 0.7)\times 10^{-10}$ \\
		B & $(6.1 \pm 0.8)\times 10^{-7}$ & 1.40 & $-1.75 \pm 0.10$ & $(3.2 \pm 0.4)\times 10^{-10}$ \\
		C & $(4.2 \pm 0.5)\times 10^{-7}$ & 1.09 & $-1.94 \pm 0.11$ & $(2.0 \pm 0.2)\times 10^{-10}$ \\
		D & $(2.2 \pm 0.2)\times 10^{-7}$ & 0.90 & $-2.14 \pm 0.08$ & $(1.05 \pm 0.08)\times 10^{-10}$ \\
		\hline
	\end{tabular}
\end{table*}
\begin{table*}
	\centering
	\caption{MAGIC + \textit{Fermi}-LAT SED fit results (shown by the dashed red line in Fig.~\ref{fig:MAGIC-LAT_SED}). Columns represent as follows: Sub-set ID; observation dates included in the sub-set; integrated flux above 80\,GeV; spectra fitting parameters for EPWL as in (\ref{eq:EPWL}); goodness of fit. Only statistical uncertainties are reported. For systematic uncertainties, please see the main text.}
	\label{tab:MAGIC-LAT_FitResults}
    \begin{tabularx}{0.9\textwidth}{ccc | c | ccclc | c}
		\hline
		Period & \multicolumn{2}{c|}{Dates} & F [cm$^{-2}$ s$^{-1}$] & \multicolumn{5}{c|}{Spectral fit parameters} & $\chi^{2}$/D.o.F\\
         & Dec. 2017 & MJD & ($E > 80$\,GeV) & $f_0$ [TeV$^{-1}$ cm$^{-2}$ s$^{-1}$] & $\alpha$ & \multicolumn{3}{c|}{$E_c$ [GeV]} & \\
		\hline
		A & 15 & 58102 & $(2.6 \pm 0.3)\times 10^{-10}$ & $(6.4 \pm 2.6)\times 10^{-8}$ & $-1.81 \pm 0.10$ & & $47\pm10$ & & $21.9/12$\\ 
        B & 16 & 58103 & $(3.1 \pm 0.3)\times 10^{-10}$ & $(7.5 \pm 2.3)\times 10^{-8}$ & $-1.72 \pm 0.08$ & & $48\pm8$ & & $25.2/17$\\
        C & 19--20 & 58106-7 & $(1.2 \pm 0.1)\times 10^{-10}$ & $(2.3 \pm 0.8)\times 10^{-8}$ & $-1.90 \pm 0.08$ & & $54\pm11$ & & $16.4/19$\\
        D & 21--29 & 58108-16 & $(2.6 \pm 0.7)\times 10^{-11}$ & $(4.3 \pm 1.7)\times 10^{-9}$ & $-2.11 \pm 0.08$ & & $59\pm17$ & & $17.4/19$\\
		\hline
	\end{tabularx}
\end{table*}

We fitted the MAGIC de-absorbed spectral points, while using the \textit{Fermi}-LAT butterfly as a constraint. The number of degrees of freedom (D.o.F) is calculated as the number of MAGIC spectral bins that contain at least one event, decreased by the number of parameters. Bins with upper limits are taken into consideration when fitting, and counted towards the number of D.o.F. The \textit{Fermi}-LAT butterfly is counted as one point. We tried fitting with power-law, power-law with exponential cut-off, log-parabola, and log-parabola with exponential cut-off. 
The best fits in all four periods were obtained with a power law with exponential cut-off (EPWL)

\begin{equation}
f(E) = f_0 \left( \frac{E}{E_0} \right)^{\alpha}\exp\left(-\frac{E}{E_c}\right),
\label{eq:EPWL}
\end{equation}
where $E_0$ is the normalization energy fixed here to 95\,GeV, $f_0$ the normalization flux, $\alpha$ the spectral slope, and $E_c$ the cut-off energy. 
The fitted SED is represented by the dashed red line in Fig.~\ref{fig:MAGIC-LAT_SED}. The fit results are given in Table~\ref{tab:MAGIC-LAT_FitResults}, where we report on statistical uncertainties only. 
The systematic uncertainty on the MAGIC energy scale is 15 per cent~\citep{MAGIC:2014zas}. 
The systematic uncertainty on the spectral slope of the MAGIC spectra in all four periods is estimated to be $\pm0.15$; and the systematic uncertainty on the cut-off energy is $15$ per cent. 
For the \textit{Fermi}-LAT data, the uncertainties were dominated by statistical ones, which were taken into account when fitting the spectra. 
Given that MAGIC systematics dominate over \textit{Fermi}-LAT ones, we estimated the resulting systematic uncertainty of the spectral slope of the joint fit at $\pm 0.10$, making the total systematic uncertainty on the slope $\pm0.18$. 

The full red line in Fig.~\ref{fig:MAGIC-LAT_SED} represents the SED on Earth and was obtained by multiplying the intrinsic SED by the gamma-ray attenuation induced by EBL. 

In all four periods, we see a smooth transition from the \textit{Fermi}-LAT to the MAGIC energy band. The resulting fit parameters are consistent in periods A and B (see Table~\ref{tab:MAGIC-LAT_FitResults}). Afterward, the \textit{Fermi}-LAT spectrum softens with a simultaneous decrease of the MAGIC flux. 
The cut-off in the spectrum occurs around 50\,GeV. 
The lowest cut-off energy occurs during the periods of highest MAGIC flux (periods A and B), and it moves towards higher energies (although remaining within uncertainties) as the flare fades out (periods C and D). 
The possible origins of the cut-off are discussed in Section~\ref{subsec:modelling}.

\subsection{\textit{Swift}}

The satellite Neil Gehrels \textit{Swift} Observatory (hereafter \textit{Swift}) is equipped with an X-ray Telescope (XRT, \citealt{Burrows2005}) and an UV/Optical Telescope (UVOT, \citealt{Roming2005}). Here we present the analysis of observations performed with UVOT and XRT. We considered observations contemporaneous to the MAGIC observations (\textit{Swift} OBSIDs: 00036381035 -- 00036381047). We also considered an extended observation period and included data between October 30 (\textit{Swift} OBSID: 00036381023) and December 31 (\textit{Swift} OBSID: 00036381048). 

\subsubsection{\textit{Swift}/UVOT}
The \textit{Swift}/UVOT telescope is a 30\,cm diffraction-limited optical-UV telescope equipped with six different filters that cover the 170--650\,nm wavelength range, in a 17\,arcmin $\times$ 17\,arcmin FoV. From the High Energy Astrophysics Science Archive Research Center (HEASARC\footnote{\href{https://heasarc.gsfc.nasa.gov/docs/archive.html}{https://heasarc.gsfc.nasa.gov/docs/archive.html}}) data base we downloaded the UVOT images in which our target sources were observed. The analysis was performed with the \verb|fappend|, \verb|uvotimsum| and \verb|uvotsource| tasks\footnote{\href{https://heasarc.gsfc.nasa.gov/docs/software/lheasoft/}{https://heasarc.gsfc.nasa.gov/docs/software/lheasoft/}}. We used a source region of 5 arcsec radius and the background was extracted from a source-free circular region with radius equal to 20\,arcsec. The extracted magnitudes were corrected for Galactic extinction using the values of \citet{Schlegel1998}, $A_B=0.072$, applying the formulae by \citet{Pei1992} for the UV filters, and finally converted into fluxes following \citet{Poole2008}. 
UVOT data are present for all four sub-sets defined from the VHE observation (periods A-D); Vega magnitudes together with statistical uncertainties\footnote{Systematic uncertainties are never greater than 0.03\,mag and the total uncertainties are therefore dominated by statistical ones} for each filter are reported in Table~\ref{tab:TableUVOT}. The UVOT data were useful to identify the low-energy peak of the spectral energy distribution and to monitor its trend during the VHE flare. In fact, as reported in Fig.~\ref{fig:MWL_SED}, the synchrotron component seems to peak at $\sim 5 \times 10^{14}$\,Hz for each period of the flare. 
The fact that the synchrotron peak frequency remains virtually constant while the flux changes is in agreement with the blazar sequence~\citep{ghisellini17}. 

\begin{table*}
\centering
\caption{\textit{Swift}/UVOT observed magnitudes. Statistical uncertainties only are reported: systematic error is always less than 0.03 mag.}
\label{tab:TableUVOT}
\begin{tabular}{l|cccccc}
\hline
Period & \textit{v} & \textit{b} & \textit{u} & \textit{w1} & \textit{m2}  & \textit{w2}\\
\hline						
A & $14.56 \pm 0.04$ & $15.00 \pm 0.03$ & $14.22 \pm 0.03$ & $14.13 \pm 0.03$ & $14.15 \pm 0.02$ & $14.25 \pm 0.03$ \\ 
B & $14.61 \pm 0.04$ & $15.05 \pm 0.03$ & $14.34 \pm 0.03$ & $14.24 \pm 0.03$ & $14.35 \pm 0.03$ & $14.47 \pm 0.03$\\
C & $15.02 \pm 0.04$ & $15.44 \pm 0.03$  & $14.69 \pm 0.03$ & $14.70 \pm 0.03$ & $14.76 \pm 0.04$ & $14.87 \pm 0.03$ \\
D & $15.75 \pm 0.06$ & $16.23 \pm 0.04$  & $15.57 \pm 0.04$ & $15.71 \pm 0.03$ & $15.57 \pm 0.04$ & $15.68  \pm 0.03$ \\
\hline
\end{tabular}
\end{table*}

\subsubsection{\textit{Swift}/XRT}
The \textit{Swift}/XRT operates in the $0.3-10$\,keV energy range. The XRT spectra were obtained from the processed event files available through the UK \textit{Swift} Science Data Centre online data products generator \citep{2009MNRAS.397.1177E}. For each observing period, time-averaged spectra were extracted and analysed using the HEASOFT v6.20 software package. The spectral files were grouped using the \texttt{grppha} task to ensure a minimum of 20 counts per bin, allowing the use of $\chi^2$ statistics. Spectral fitting was performed with XSPEC v12.9.1 \citep{Dorman2001}. 
The spectra were modeled with an absorbed power-law model. Photoelectric absorption was taken into account using a fixed hydrogen column density corresponding to the Galactic value along the line of sight, $N_\text{H}=1.77 \times 10^{20}$\,cm$^{-2}$~\citep{2013MNRAS.431..394W}. For each period, the absorbed power-law model provides a good description of the data. From the spectral fits, we derived the photon index $\alpha$ and the unabsorbed flux in the $0.3-10$\,keV energy band. The best-fit spectral parameters are reported in Table~\ref{tab:TableXRT}. 
The flux appears to be rather high, but constant at the beginning of the extended observation period. Between MJD 58056 and 58076, the average flux value is about 20 per cent higher than the highest flux observed during the MAGIC flare. This coincides partially with the highest flux in \textit{Fermi}-LAT, although because of the sparse X-ray sampling and the lack of observations in other wavebands, it is impossible to determine the origin of the high X-ray flux and whether it follows the variability visible in the HE gamma-ray band. Later, during the MAGIC flare, the highest X-ray flux is lower by some 25 per cent, compared to the highest flux overall, and it decays towards the end of the MAGIC flare. 
The X-ray spectrum plays a fundamental role in constraining the emission models, since the X-ray region corresponds to the valley between the synchrotron and the high-energy bump. 
As can be seen in Fig.~\ref{fig:MWL_SED}, the XRT flux contributes to the high-energy peak. 

\begin{table}
\footnotesize
\centering
\caption{Results of the \textit{Swift}/XRT data analysis.}
\label{tab:TableXRT}
\begin{tabular}{l|cccc}
\hline
Period & Exp. time & $\alpha$ &  $\chi^2_{\textit{red}} ($d.o.f.)  & F$_{0.3-10 \text{keV}}$\\
       & [ks]      &           &                                   & [$10^{-12}$\,erg\,cm$^{-2}$\,s$^{-1}$]\\
\hline
A & 1.987 & 1.60 & 0.58/4 & 5.165\\
B & 0.981 & 1.44 & 1.66/1 & 7.057 \\
C & 2.804 & 1.60 & 0.95/6 & 4.068 \\
D & 4.428 & 1.56 & 0.50/5 & 2.662\\
\hline
\end{tabular}
\end{table}

\subsection{WEBT}
\label{subsec:webt} 

TON\,0599 is one of the sources regularly monitored by the GASP of the WEBT\footnote{\url{https://www.oato.inaf.it/blazars/webt/}} \citep[e.g.][]{villata2002,villata2009,raiteri2013,raiteri2017,jorstad2022}.
The optical data provided by the GASP-WEBT observers for this work were acquired at the following observatories: 
Abastumani (Georgia),                
Crimean (Crimea), 
Lowell (USA; Perkins and DCT telescopes),
Lulin (Taiwan),                          
Mt. Maidanak (Uzbekistan; 60, 100, and 150 cm telescopes),            
New Mexico Skies (USA),                     
Roque de los Muchachos (Spain; TNG, NOT, and Liverpool telescopes),                                             
Rozhen (Bulgaria; 50/70 cm telescope),                   
Sirio (Italy),                                
St. Petersburg (Russia),                       
Teide (Spain; IAC80 and STELLA-I telescopes),            
Tien Shan (Kazakhstan),                           
Tijarafe (Spain),
Vidojevica (Serbia). 
Further optical data were obtained through the telescope network of the Las Cumbres Observatory at 
Cerro Tololo (Chile),                   
Haleakala (USA),                      
McDonald (USA),                       
Teide (Spain).                         

The source magnitude  was calibrated using the photometric sequence by \citet{rai98} in the Johnson-Cousins $B$, $V$, and $R$ bands, and according to \citet{gon01} in the $I$ band.

NOT observations were performed in the Sloan Digital Sky Survey (SDSS) $ugriz$ filters; they were converted into Johnson-Cousins values with the transformations by \citet{jor06}. The Liverpool Telescope data were acquired with the `blue', `green' and `red' cameras of the RINGO3 optical imaging polarimeter. They were rescaled to match the $V$, $R$, and $I$ Johnson-Cousins bands, respectively, through comparison with the other datasets.

The WEBT datasets were complemented by $V$ and $R$ band data from the ground-based observational support program to the {\it Fermi} satellite running at the Steward Observatory\footnote{\url{https://james.as.arizona.edu/~psmith/Fermi/}} \citep{smith2009}.

\begin{figure}
	\includegraphics[width=\columnwidth]{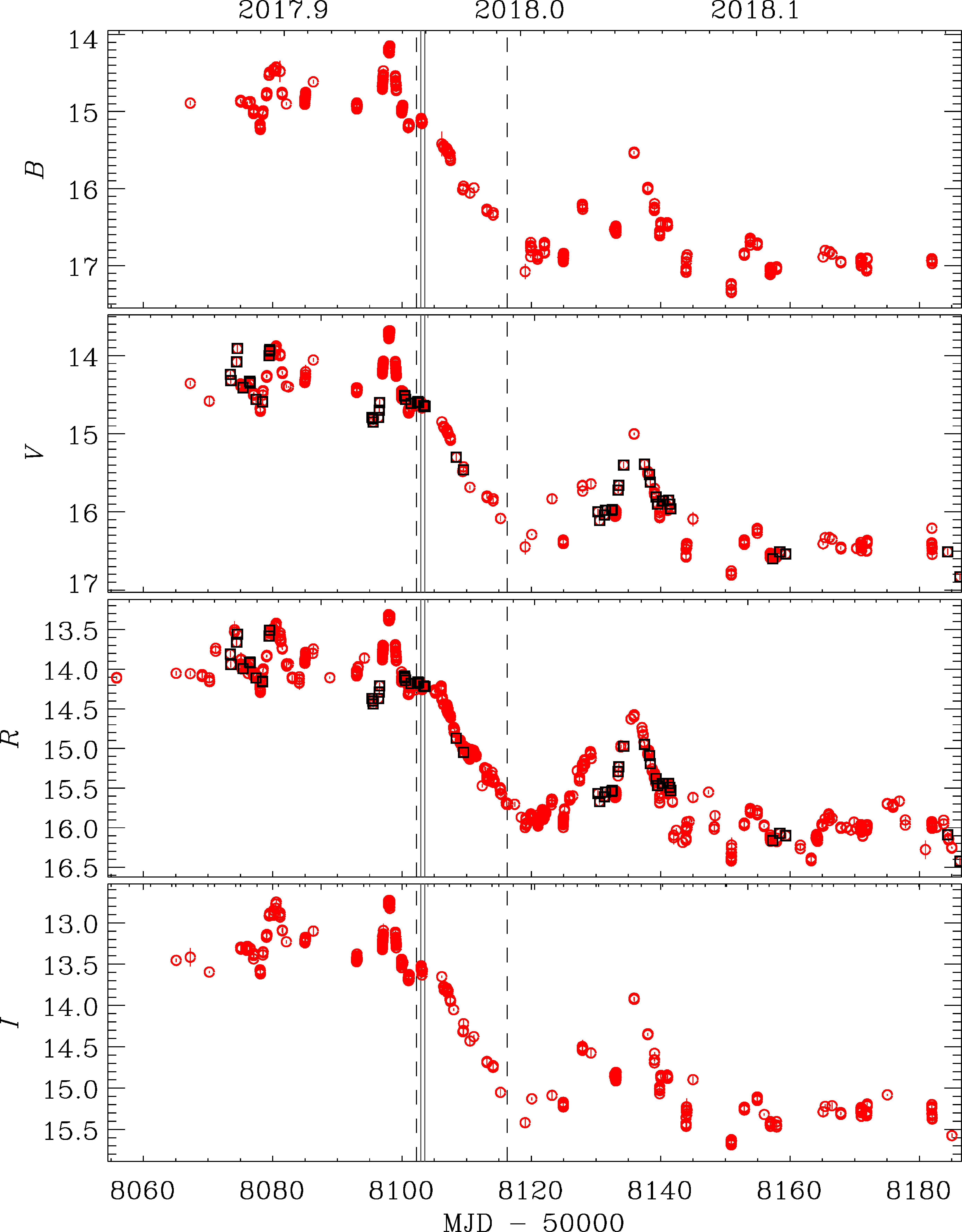}
    \caption{From top to bottom: $B$, $V$, $R$, $I$ light curves (observed magnitudes) obtained with data from the WEBT collaboration (red circles) and Steward Observatory (black squares). Vertical dashed lines mark the beginning and the end of the MAGIC observations, vertical full lines mark the {\it Swift} pointings of 2017 December 15 (Period A) and 16 (Period B).}
    \label{fig:webt}
\end{figure}

The optical light curves from November 2017 to March 2018 are shown in Fig.~\ref{fig:webt} as observed magnitudes and in Fig.~\ref{fig:LC_MWL} the best-sampled $R$ band light curve is plotted as dereddened flux densities, after correction for a Galactic extinction of 0.043\,mag. 
Light curves were processed in order to identify and correct for offsets between different datasets and to remove bad points, i.e. strong outliers or points affected by large uncertainties (greater than 0.2\,mag).

Optical polarimetric data were taken at the Crimean, Lowell (Perkins), Roque (Liverpool Telescope), St. Petersburg, and Steward observatories. Fig.~\ref{fig:LC_MWL} shows the behaviour in time of the polarization degree and electric vector polarization angle (EVPA). 

Radio observations were collected as part of the WEBT campaign. They were performed at the Noto and Medicina \citep[Italy,][]{Giroletti:2020olc}, Mets\"ahovi \citep[Finland,][]{1998A&AS..132..305T}, Pico Veleta (Spain, IRAM 30~m telescope) observatories and at the Submillimeter Array (SMA, Maunakea, USA). 
Measurements from the IRAM 30~m Telescope were obtained as part of the POLAMI (Polarimetric Monitoring of AGN at Millimetre Wavelengths) Program\footnote{\url{http://polami.iaa.es}} \citep{agudo2018a,agudo2018b,thum2018}.
The SMA data are from the SMA Calibrator Flux Density database\footnote{\url{http://sma1.sma.hawaii.edu/callist/callist.html}} \citep{gurwell2007}.
The radio light curves are shown in Fig.~\ref{fig:LC_MWL}.

The flux in the $R$ band shows the highest variability during the extended observation period. It is more than 20 percent more variable than the \textit{Fermi}-LAT flux and more than 40 per cent more variable than the X-ray flux in the same period (see Section~\ref{subsec:Fvar} for more details). However, the period that contributes the most to the flux variability is before the MAGIC observation window. 
In particular, there are three flares that occurred in the MJD periods $58073.5 - 58074.5$, $58078.0 - 58084.0$, and $58095.0 - 58100.0$. To determine the variability timescale, we fitted all three features with a Gaussian. The fit results for the three flares are: $\mu_1 = 58074.11 \pm 0.01$\,MJD, $\sigma_1 = 0.64 \pm 0.02$ days, $\mu_2 = 58080.50 \pm 0.03$\,MJD, $\sigma_2 = 2.04 \pm 0.04$ days, and $\mu_3 = 58098.14 \pm 0.02$\,MJD, $\sigma_3 = 1.47 \pm 0.04$ days. Observations with the \textit{Swift} satellite are not dense enough to make any connections to the emission in UV and X-ray bands. However, the light curve in the HE gamma-ray band shows no prominent flares, indicating that the flares in the $R$ band were not fully connected to the gamma-ray emission region.

\subsection{Optical spectroscopy}
An optical spectrum of TON\,0599 was acquired at the 3.6-m Telescopio Nazionale Galileo (TNG) on 2018 February 22 with the Dolores instrument. We used a 1.5\,arcsec slit and the LR-B grism. The exposure time was 1200\,s. 

\begin{figure}
	\includegraphics[width=\columnwidth]{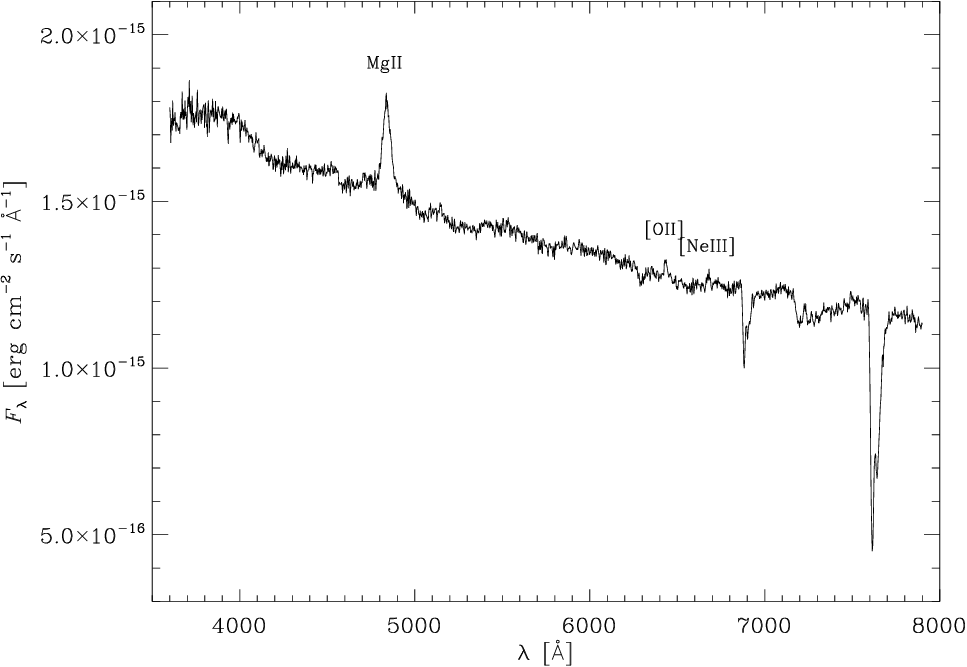}
    \caption{Optical spectrum of TON\,0599 acquired at TNG on 2018 February 22 with the Dolores instrument.}
    \label{fig:spot}
\end{figure}

The spectrum of TON~0599 was flux calibrated with a spectrum of the spectrophotometric standard star Feige~66 using IRAF's standard tasks (precisely, the tasks standard, sensfunc, and calibrate in the onedspec package) and the extinction values for La Palma computed by~\citet{King85}. 
The calibrated spectrum is shown in Fig.~\ref{fig:spot}. 
It shows a prominent MgII $\lambda 2800$ broad emission line, with a line flux of $\sim 1.6 \times  10^{-14} \rm \, erg \, cm^{-2} \, s^{-1}$ and a $\rm FWHM \sim 51$ \AA, which implies a velocity of $\sim 3200 \rm \, km \, s^{-1}$. 
This result is consistent with the result of \citet{2022ApJ...926..180H} for the same period. 
Other visible features are the narrow emission lines of [OII] $\lambda$3727 and [NeIII] $\lambda$3869 with fluxes of $1.3 \times  10^{-15} \rm \, erg \, cm^{-2} \, s^{-1}$ and $\sim 1.0 \times  10^{-15} \rm \, erg \, cm^{-2} \, s^{-1}$, respectively. From their $\rm FWHM \sim 20$ \AA\ one can derive a velocity of about $1100 \rm \, km \, s^{-1}$.
We notice that the flux ratios MgII/[OIII] and MgII/[NeIII] are about twice smaller than in quasars \citep{van01}.
Full results are given in Table~\ref{tab:opticalspectrum}. The [NeIII] measurements are very challenging due to the quality of the spectrum. Therefore, these values are quoted as approximate and omitted from the table.

\begin{table}
	\centering
 	\caption{Properties of the optical emission lines detected from the TON\,0599 spectrum shown in Fig.~\ref{fig:spot}.}
	\label{tab:opticalspectrum}
    \begin{tabular}{l|c|c}
Line ID & MgII 2800 & [OII] 3727 \\
EW (\AA) & 10.2 & 1.0 \\
FWHM (\AA) & 51 & 23 \\
FWHM (km/s) & 3170 & 1080 \\
Flux (erg/cm$^2$/sec) & $1.6\times 10^{-14}$ & $1.3\times 10^{-15}$ \\ 
Luminosity (erg/s) & $3.8\times 10^{43}$  & $3.1\times 10^{42}$ 
\end{tabular}
\end{table}

\section{Multiwavelength picture}
\label{sec:MWL}
\subsection{Multiwavelength light curve}
\label{subsec:MWL_LC}
\begin{figure*}
    \centering
	\includegraphics[width=2\columnwidth]{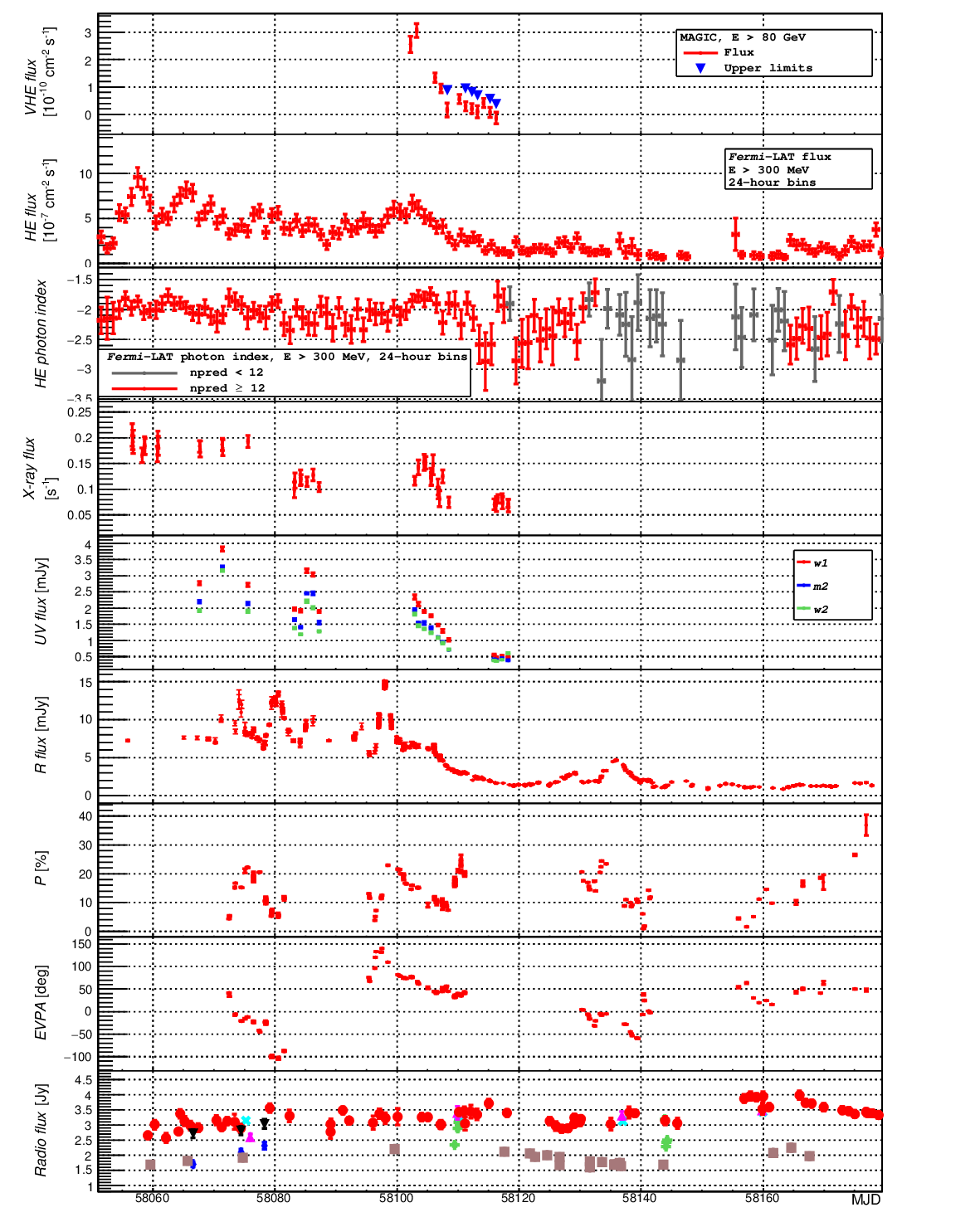}
    \caption{Multiwavelength light curve. Data shown in pads from top to bottom: i) Flux above 80\,GeV measured by MAGIC (same data as in Fig.~\ref{fig:LC_MAGIC}), blue triangles represent upper limits for fluxes whose relative uncertainty is greater than 0.5. ii) Flux above 300\,MeV measured by \textit{Fermi}-LAT. iii) Spectral photon index for \textit{Fermi}-LAT measurements. \textit{Fermi}-LAT points are based on 24 hour integration. Points with npred\,$\geq12$ are shown in red. iv) X-ray photon flux measured with \textit{Swift}/XRT. v) UV fluxes with \textit{w1}, \textit{m2} and \textit{w2} filters measured with \textit{Swift}/UVOT. vi) Optical flux densities in the \textit{R} band. They have been corrected for the Galactic extinction. 
    vii) Optical polarization degree. viii) Optical polarization angle. ix) Radio fluxes at 4.8\,GHz from Noto (green plus signs), 8.5 GHz from Medicina (cyan crosses), 24 GHz from Medicina (magenta triangles), 37 GHz from Mets\"ahovi (red circles), 86 GHz from Pico Veleta (black triangles), and 229 GHz from Pico Veleta (blue diamonds) and Maunakea (brown squares). 
    }
    \label{fig:LC_MWL}
\end{figure*}

We constructed a MWL light curve in Fig.~\ref{fig:LC_MWL}, presenting the flux above 80\,GeV measured with the MAGIC telescopes (also shown in Fig.~\ref{fig:LC_MAGIC}); flux and photon index in the 0.3-500\,GeV energy range measured with \textit{Fermi}-LAT, X-ray and 
optical--UV flux measured by XRT and UVOT instruments (respectively) on board the \textit{Swift} satellite; optical flux densities in 
$R$ band corrected for Galactic extinction, obtained by the WEBT collaboration and Steward Observatory; degree and angle of optical polarization; and radio flux densities measured at 4.8, 8.5, 24, 37, 86 and 229 GHz. In order to study the evolution of the flux, we extended the time period beyond the MAGIC observations and studied fluxes from 2017 October 25 (MJD 58051) to 2018 March 2 (MJD 58179).

\begin{figure}
	\includegraphics[width=\columnwidth]{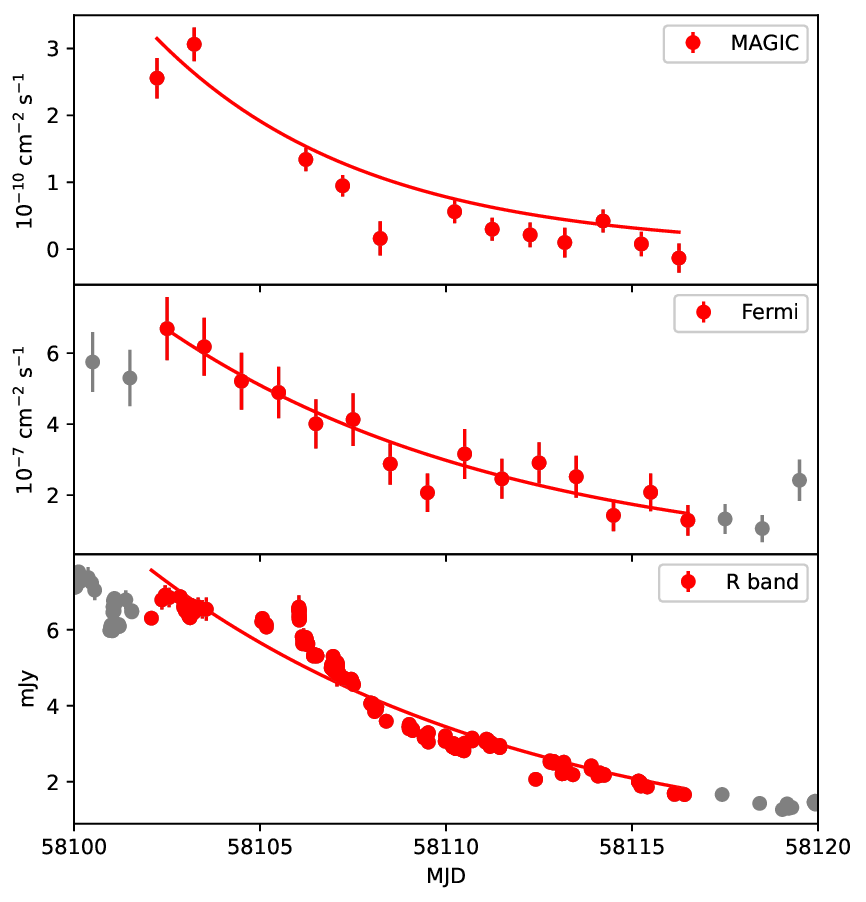}
    \caption{MWL light curve around the MAGIC observations fitted with exponential function.}
    \label{fig:decay}
\end{figure}

There is some noticeable similarity between the light curves in different wavelengths, particularly during the MAGIC observation window. A change of flux simultaneous to MAGIC is evident in all wavelengths except the radio bands, where we see no significant change of flux. 
In order to compare the decay time of the fluxes in different bands, we fit fluxes in VHE, HE, and $R$ bands during the MAGIC observation window with a decay exponential function 

\begin{equation}
f(t) = f_0 \exp\left( -(t - t_0)/\tau\right),
\label{eq:decay}
\end{equation}
where the parameter $t_0$ was fixed to the beginning of the MAGIC observations (MJD 58102.23). 
The light curves in the other bands were not considered because of a hole in the dataset of almost 3 days within the given period. 
In Table~\ref{tab:decay}, we give the flux halving time $\tau_{1/2} = \tau\ln(2)$, and the reduced $\chi^2$ value. 
The MWL light curve for this period is shown in Fig.~\ref{fig:decay}. 
The VHE band experiences the fastest decay with a halving time of $3.9 \pm 0.9$ days, whereas the HE and \textit{R}-band fluxes decay at a comparable rate, slower than the VHE flux. 
The $R$ band points have much smaller uncertainties, leading to a much worse $\chi^2$ for the fit. 

\begin{table}
	\centering
 	\caption{Properties of the decay rates of fluxes in different bands during the MAGIC observation window. $\tau_{1/2}$ represents the time needed for flux to decay to half of its starting value. D.o.F. stand for degree of freedom.}
	\label{tab:decay}
    \begin{tabular}{l|c|l}
        Band & $\tau_{1/2}$ [day] & $\chi^2 / \mathrm{D.o.F.}$\\
        \hline
        VHE & $3.9 \pm 0.9$ & ~4.5 \\
        HE & $6.5 \pm 0.5$ & ~0.78 \\
        R band & $7.0 \pm 0.1$ & 78 \\
    \end{tabular}
\end{table}

On the other hand, there are some obvious differences between the light curves in different wavebands. The HE flare happening around  MJD 58057.6 does not have an obvious counterpart. Admittedly, the same time period was covered only in X-rays and only sparsely, where the flux is highest in the whole observation window but apparently constant. 
Another example of outstanding features are relatively sharp peaks in the $R$ band before the MAGIC observation window. Again, we miss the data for the same time period in most of the bands; however, there were simultaneous observations with \textit{Fermi}-LAT that show no prominent simultaneous flares. 

In order to further investigate the connection between different bands, we calculated the Spearman's rank coefficient of correlation between them. Using the code \textsc{matchLCs}\footnote{\href{https://github.com/tterzic/matchLCs}{https://github.com/tterzic/matchLCs}}, we matched closest measurements in different bands by requiring that two measurements in different bands were performed within a certain time window. Observations with MAGIC, \textit{Fermi}-LAT or \textit{Swift}/UVOT were performed with a $\sim 1$\,day cadence. Therefore, when one of these bands was considered, the simultaneity window was set to $\pm 12$ hours. In other cases, the sampling was denser than one measurement per day, so we requested the time difference of up to $\pm 6$ hours in order for a pair to be considered simultaneous. 
When calculating correlation with the \textit{Fermi}-LAT photon index, only points with npred\,$\ge$\,12 (red points in the third panel in Fig.~\ref{fig:LC_MWL}) were considered. 

Flux measurement uncertainties were taken into account by simulating $10^6$ points for each measured point, assuming a Gaussian distribution with the standard deviation corresponding to the measurement uncertainty and no correlation between bands or measurements in a given band. Then the Spearman's rank correlation coefficient was calculated for each of the $10^6$ simulated datasets. In Table~\ref{tab:Corrleation}, we give the mean correlation coefficient, the standard deviation on the last digit in round brackets, and the number of pairs between bands in square brackets.
One should note that simulating points while assuming no correlation is reflected on their distributions and, consequently, on the individual, as well as the mean correlation coefficient, which results in somewhat underestimated values of correlation.
\begin{table*}
	\centering
 	\caption{Spearman's correlation coefficient between different bands based on $10^6$ simulations per band pair. The numbers in round brackets give standard deviation on the last digit, while the number of pairs on which the correlation was calculated is given in square brackets.}
	\label{tab:Corrleation}
    \begin{tabular}{l|ccccccc}
\hline
Band & \multicolumn{7}{c}{Spearman's rank correlation coefficient}  \\
\hline
P\,\% & $0.2(2)\,[13]$ \\
R-band & $-0.43(7)\,[36]$ & $-0.127(8)\,[127]$ \\
UVw1 & $-0.5(2)\,[7]$ & $0.99(2)\,[6]$ & $0.961(9)\,[16]$ \\
XRT & $-0.6(3)\,[4]$ & $0.9(1)\,[6]$ & $0.71(7)\,[16]$ & $0.74(6)\,[19]$ \\
LAT index &  $-0.3(1)\,[44]$ & $-0.3(1)\,[32]$ & $0.25(9)\,[69]$ & $-0.2(2)\,[18]$ & $-0.1(2)\,[21]$ \\
LAT flux &  $-0.42(7)\,[53]$ & $0.07(5)\,[43]$ & $0.78(2)\,[86]$ & $0.66(8)\,[19]$ & $0.67(7)\,[22]$ & $0.24(8)\,[114]$ \\
MAGIC & $-0.4(3)\,[6]$ & $0.1(1)\,[7]$ & $0.81(8)\,[12]$ & $0.98(5)\,[5]$ & $0.8(2)\,[5]$ & $0.3(2)\,[12]$ & $0.7(1)\,[12]$ \\
\hline
 & 37\,GHz & P\,\% & R-band & UVw1 & XRT & LAT index & LAT flux \\
\hline \hline
\end{tabular}
\end{table*}

The Spearman's rank correlation coefficient between fluxes in the VHE and HE bands was found to be expectedly strong: 0.7, based on 12 pairs. 
Visual inspection shows very similar trends in the VHE and X-ray or UV bands. Indeed, the correlation coefficients are 0.98 and 0.8, respectively, however, in both cases, there were only 5 pairs of measurements between MAGIC and \textit{Swift}. 
It is interesting to note that the correlation of the VHE flux with the optical $R$ band is 0.81 on 12 pairs. The flux in the $R$ band shows quite some variability in the period prior to the MAGIC observations. Comparing the \textit{Fermi}-LAT and the $R$ band light curves on the whole period presented in Fig.~\ref{fig:LC_MWL} results in a correlation coefficient of 0.78 based on 86 simultaneous observations. Such simultaneous variations of fluxes over a longer time period indicate the same emission region for these bands. 
There is a significant correlation between X-ray and UV with correlation coefficients of 0.74, based on 19 pairs. 
The degree of the optical polarization shows a strong correlation with X-ray and UV fluxes, 
but based on 6 pairs only. 
Other bands do not seem to be correlated with the degree of polarization. As we will see in Section~\ref{subsec:Fvar}, the variability of polarization is not negligible, however, it varies with a different pattern compared to flux in different bands (see Fig.~\ref{fig:LC_MWL}). 
It is curious to note that although the flux in the gamma rays and optical band stays rather low after the flare observed by MAGIC (past MJD $\sim58120$) until the end of the extended observation period, there is a substantial increase in the degree of polarization, which reaches 40 per cent (within uncertainties) at the end of the extended period. 
Unfortunately, polarization data are rather sparse, hindering a more detailed investigation. 
In radio, we inspected the 37\,GHz band because it is the most densely sampled one, however, the number of matching observations with some bands is still rather small. Nevertheless, it is interesting to point out that all bands show a moderate negative correlation with the radio band. Exceptions are the degree of optical polarization and the \textit{Fermi}-LAT photon index, both of which show low degrees of positive or negative correlation, respectively.

\subsection{Fractional variability}
\label{subsec:Fvar}
\begin{figure}
    \includegraphics[width=0.95\columnwidth]{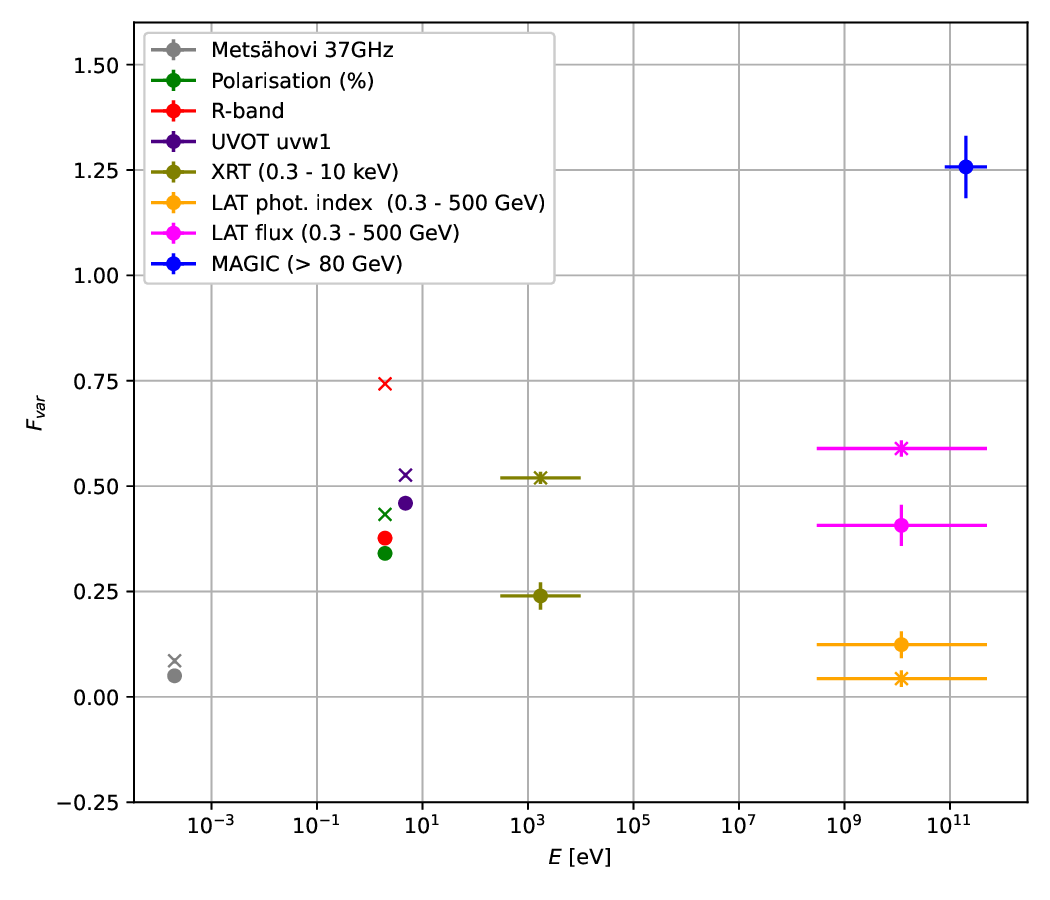}
    \caption{Fractional variability of TON\,0599 in different bands. Circles represent measurements during the MAGIC observation window (MJD 58102 -- 58116), while results marked with crosses were obtained for the whole period presented in Fig.~\ref{fig:LC_MWL} (MJD 58051 -- 58180). Horizontal error bars represent the respective energy bands.}
    \label{fig:Fvar}
\end{figure}

In order to compare the variability in different bands, we calculate the fractional variability ($F_\mathrm{var}$) for all the data shown in Fig.~\ref{fig:LC_MWL}. 
$F_\mathrm{var}$ is defined as the relative root mean square of the intrinsic variability corrected for the effect of random errors
\citep{Rodriguez-Pascual_1997}:

\begin{equation}
  F_\mathrm{var}= \sqrt{\dfrac{S^{2} -\langle\sigma_\mathrm{err}^{2}\rangle}{\langle F_{\gamma} \rangle^{2}}},
\end{equation}\label{eq:fvar}
where $S$ is the standard deviation of $N$ measurements,
$\langle\sigma_\mathrm{err}^{2}\rangle$ is the mean squared error, and
$\langle F_{\gamma}\rangle$ is the average measurement value. 
In case the variation of values is dominated by the measurement uncertainties, the fractional variability is calculated as 

\begin{equation}
  F_\mathrm{var}= -\sqrt{\dfrac{\langle\sigma_\mathrm{err}^{2}\rangle - S^{2}}{\langle F_{\gamma} \rangle^{2}}}.
\end{equation}\label{eq:fvarneg}
Note that in this case $F_\mathrm{var}$ is defined as negative.
The uncertainty is estimated following the recipe from \citet{2008MNRAS.389.1427P}:

\begin{equation}
  \Delta F_\mathrm{var} = \sqrt{F_\mathrm{var}^2 + err(\sigma_\mathrm{NXS}^2)}-F_\mathrm{var},
\end{equation}
where the error on the normalized excess variance \citep{2003MNRAS.345.1271V} is given as 

\begin{equation}
  err\left(\sigma_\mathrm{NXS}^2\right)=\sqrt{ \left( \sqrt{\frac{2}{N}} \cdot \frac{\langle \sigma_\mathrm{err}^{2}\rangle}{\langle F_{\gamma} \rangle^{2}} \right)^{2} + \left(\sqrt{\frac{\langle\sigma_\mathrm{err}^{2}\rangle}{N}} \cdot \frac{2 F_\mathrm{var}}{\langle F_{\gamma}\rangle} \right)^{2}}.
\end{equation}
We calculated $F_\mathrm{var}$ for i) the whole period presented in Fig.~\ref{fig:LC_MWL}, and ii) the MAGIC observation window. Results are shown in Fig.~\ref{fig:Fvar}~\footnote{Strong variability of EVPA is apparent from Fig.~\ref{fig:LC_MWL}. However, showing its $F_\mathrm{var}$ would completely dominate Fig.~\ref{fig:Fvar}. In addition, the calculation of $F_\mathrm{var}$ would not be clear because of the 180\,deg EVPA ambiguity. For these reasons we omitted $F_\mathrm{var}$ for EVPA.}. As in the correlation study, only \textit{Fermi}-LAT photon indices with npred\,$\ge$\,12 were used. 
Since the X-ray region corresponds to the valley between the synchrotron peak and the high energy bump (see Section~\ref{subsec:modelling}), its fractional variability is somewhat smaller than the variability in optical and UV bands, which correspond to the synchrotron peak frequency. 
Optical and X-ray variabilities for the whole observation period are double than the respective values during the MAGIC observation window, because of large differences between the high flux before and the low flux after the MAGIC observations. 
It is interesting to notice how in gamma rays the variability is much higher for MAGIC data, than for \textit{Fermi}-LAT. Looking at the gamma-ray spectral evolution in Fig.~\ref{fig:MAGIC-LAT_SED}, one can see how a small change of spectral slope below 10\,GeV can induce a large change of flux above a few tens of GeV even without the change in the position of the cut-off. 
In almost all cases, the variability is greater for the longer period than for the duration of the MAGIC flare. An exception is the \textit{Fermi}-LAT photon index. Its low $F_\mathrm{var}$ results from large measurement uncertainties, which are comparable to the spread of the data. As expected from the MWL light curve, almost no variability is seen at 37\,GHz. 

\citet{Schleicher:2019lkf} discussed how the completeness of datasets affects fractional variability and found that the relative systematic uncertainties due to sampling can be as large as 10 percent. This could particularly influence the comparison between the variabilities in the two time intervals. However, in our case, the higher variability for the longer period is most likely due to large differences in fluxes over a longer time interval. Moreover, as \citet{Schleicher:2019lkf} themselves argue, the systematic uncertainties are too small to affect the conclusions on the
general trends.

\subsection{SED modelling}
\label{subsec:modelling}
We model the emission from TON\,0599 using the simple one-zone leptonic model fully described in \cite{Maraschi2003}. In this scheme, the low-energy peak is modeled as synchrotron emission from relativistic electrons carried by the jet. The high-energy peak is attributed to IC from the same electrons \citep{Ghisellini1998}. In FSRQs the dominant target photon population is generally assumed to be external, i.e. the radiation from the broad line region (BLR) and/or the dusty torus (e.g. \citealt{gh09}). We consider a homogeneous blob of relativistic electrons moving down the jet with bulk Lorentz factor $\Gamma_b$ and carrying a tangled magnetic field with (comoving) strength $B^{\prime}$. In the frame of the emission region the electrons follow a smoothed broken power-law energy distribution extending from $\gamma^{\prime}_{\mathrm{min}}$ to $\gamma^{\prime}_{\mathrm{max}}$. Spectral indices below and above the break at $\gamma^{\prime}_{\mathrm{brk}}$ are $p_{1}$ and $p_{2}$, respectively, and the normalization is $K^{\prime}$ (all primed quantities are expressed in the source frame). We model emission in each period separately and independently. We consider the emission in each period as static, i.e. we do not model transition from one period to the next.

According to \citet{gh09} the radii of the BLR and dust torus are proportional to the square root of total emission of the accretion disk. The disk luminosity is estimated from the BLR emission lines luminosity based on a recipe from \citet{gh15} to be $3.6\times 10^{45}$\,erg\,s$^{-1}$, which is similar, although below the lower limit used in \citet{patel2018}. This gives $1.9\times 10^{17}$\,cm and $4.7\times 10^{18}$\,cm for the radii of the BLR and dust torus, respectively. 
Given the prominent broad line emission, a strong absorption of gamma rays would be expected if the gamma-ray emission zone was within the BLR zone (see e.g. \citealt{donea2003,sitarek2008,tavecchio2009,poutanen2010,Costamante2018}). Indeed we do detect a cut-off in the spectra between \textit{Fermi}-LAT and MAGIC bands (see Table~\ref{tab:MAGIC-LAT_FitResults}), however at energies significantly higher than the expected $\sim20$\,GeV, and the $\sim12$\,GeV reported by \citet{10.1093/mnras/stz3490}. Therefore, as suggested for the case of other FSRQ showing the same behaviour (see e.g. \citealt{Aleksic2011, Aleksic2014, ahnen2015, ahnen2016}), we place the emission region outside of the BLR, setting a lower limit on the emission zone distance from the black hole (BH) $d>1.9\times 10^{17}$\,cm. This estimate of the BLR radius is consistent with the values previously reported in \citeauthor{patel2018} (\citeyear{patel2018}, $(2.11-2.45)\times 10^{17}$\,cm) or \citeauthor{prince2019} (\citeyear{prince2019}, $(2.4-2.98)\times 10^{17}$\,cm). 
Assuming the jet aperture angle $\theta_{j}=0.1$\,rad, one obtains a lower limit on the radius of the emission zone of $R>1.9\times10^{16}$\,cm. 
We set the upper limits on the gamma-ray emission zone based on the causality argument

\begin{equation}
    R \le c t_\mathrm{var} \frac{\delta}{1+z},
\end{equation}
where $c$ is the speed of light, $\delta$ is the relativistic Doppler factor, and $z$ redshift of the source. For the variability timescale $t_\mathrm{var}$, we take the flux halving time from Table~\ref{tab:decay}. Using $t_\mathrm{var}=3.9$\,days and $\delta=20$ (intermediate value from our modelling reported in Table~\ref{tab:SED_modelParameters}), we obtain $R\le 1.2 \times 10^{17}$\,cm. 
The final emission zone radius is obtained from the model fit, and the distance from the BH is fixed at 10 times the value. By construction, the only important external radiation field is the one produced by the torus (see e.g. \citealt{gh09}), modelled as a black body at 1000\,K. 

\begin{figure*}
	\includegraphics[width=\textwidth]{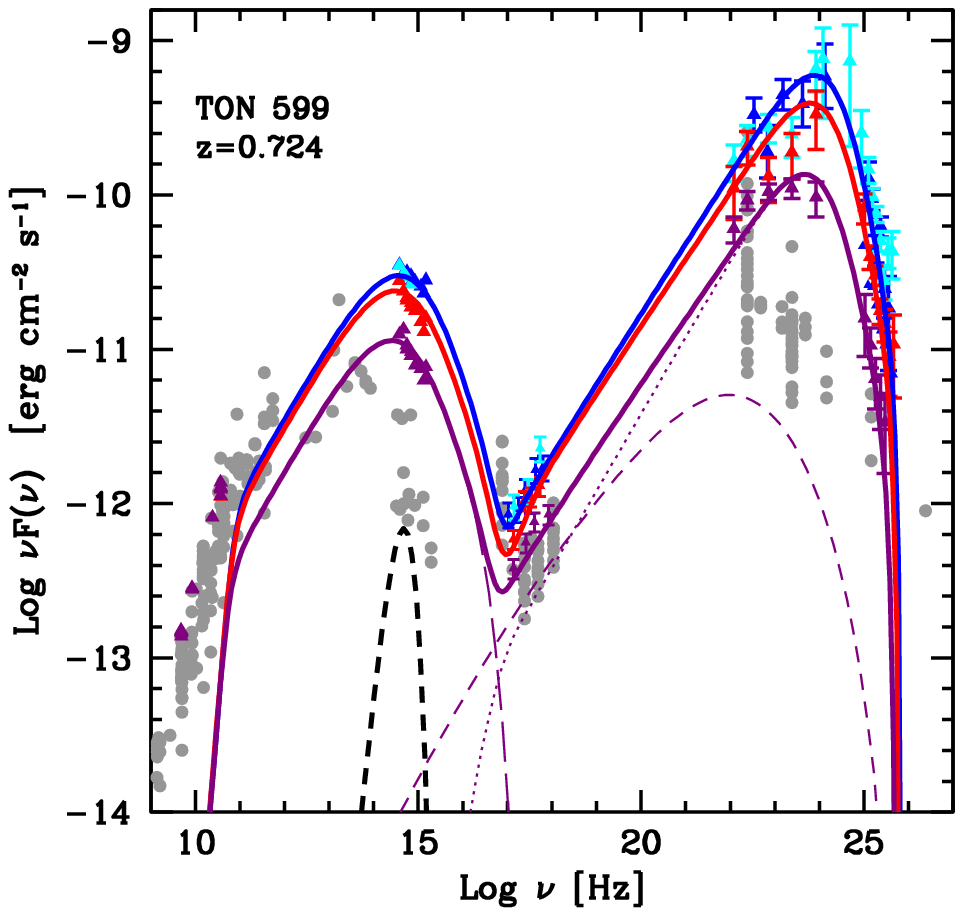}
    \caption{Spectral energy distribution for TON\,0599. Data from each period and respective models are colour coded as follows: Period A -- blue, Period B -- light blue, Period C -- red, Period D -- purple. Black dashed curve represents the dust torus. Grey points are historical measurements. For state D we also separately report the contribution from the different emission components: synchrotron -- long-dashed, SSC -- short-dashed and EC -- dotted.}
    \label{fig:MWL_SED}
\end{figure*}

The results of the modelling are shown in Fig.~\ref{fig:MWL_SED}, and the inferred model parameters are reported in Table~\ref{tab:SED_modelParameters}. The parameters are found by judging ``by eye'' the agreement between the model output and the observational data points. 
Moreover, we tried to minimize the variations of the parameters between the different states, and the two slopes of the electron energy distribution have been fixed.
Since the models corresponding to states A and B are very similar, in Fig.~\ref{fig:MWL_SED} and Table~\ref{tab:SED_modelParameters} we only report the model and the parameters corresponding to state A. 
The values that we got for some of the parameters are substantially different from the ones reported in \citet{patel2018} (e.g. we got a factor of $\sim10-30$ larger emission zone radius, while a factor of $\sim30-80$ weaker magnetic field); however, \citet{patel2018} base their estimates on data in X-ray and lower energy bands and fit the SED with a second degree polynomial, while our results are based on broadband modelling across the whole electromagnetic spectrum and using a physical scenario. In our model synchrotron radiation accounts for the low-energy bump, up to the `valley' between the two bumps. 
synchrotron self-Compton (SSC) contributes substantially (about 50 per cent) only in the X-ray band (see Fig.~\ref{fig:MWL_SED}, where we separately report the contributions of the different emission components for state D). Above the X-ray band, the emission is dominated by external Compton (EC) scattering on the torus photons. The different states are reproduced with slight variations of some of the model parameters, while the main difference between states is a result of the variation of the bulk Lorentz factor, which decreases with time. However, in view of the degeneracies characterising the emission model, it is possible that other combinations of the parameters can satisfactorily reproduce the data. 

We remark that in addition to the distance to the BH (which with the standard assumption $\theta_j=0.1$ also determines the size of the emission region), we also constrain the level of the external radiation field (albeit by using a standard scaling relation between radius and luminosity). This is enough to fix the magnetic field and, in turn, the energy of the particles emitting at the peak of the SED. We also stress that the degeneracy between some of the parameters, coupled with the quality and the spectral coverage of the data, imply non-negligible uncertainties for the inferred parameters.

Considering $\Gamma _b \ge 10$ (as we get from our emission modelling consistently with standard modelling, e.g. \citealt{gh09}), one would expect that the blob would travel for large distances on a time-scale of 10 days. However, a presence of a stationary (in observer's frame) shock, in which the plasma flows at relativistic speed (see e.g. \citealt{bodo2018}) could explain high bulk Lorentz factors without the emission region covering large distances. 
The same scenario would explain lower flux levels corresponding to lower values of $\Gamma_b$.

This scenario does not exclude some degree of dynamical behaviour. For instance, \cite{bodo2018} studied the effect of a density perturbation on the structure of the recollimation shock and found that the structure of the shock (that, in turn, influences the value of physical quantities in the downstream region, where the emission occurs) shows important variations. In particular, the change of the shock shape (and therefore of the inclination of the upstream velocity with respect to the shock front) determines strong variations of the downstream velocity pattern (both in direction and intensity)  that can strongly affect the inferred effective Lorentz factor and size used in our simplified one-zone model.
A detailed study of the feasibility of the proposed scenario is beyond the scope  of the present article. 

The cut-off present in the \textit{Fermi}-LAT -- MAGIC spectrum can be explained by the combination of the maximum energy of relativistic electrons and the decline of the IC scattering efficiency in the Klein-Nishina regime, where the latter effect is the dominant one.

In the simple framework adopted here (one-zone, tangled magnetic field), the temporal modelling of the polarization is not included. A proper modelling of the highly structured variability of optical polarization (degree and angle) displayed in Fig.~\ref{fig:LC_MWL} seems however a quite difficult task for current standard models. 

\begin{table*}
	\centering
 	\caption{Parameters of the SED models shown in Fig.~\ref{fig:MWL_SED}.}
	\label{tab:SED_modelParameters}
    \begin{tabular}{lccccc}
\hline
Parameter & Symbol & \multicolumn{3}{c}{Period} \\
 & & A / B & C & D \\
\hline \hline
Minimum Electron Lorentz Factor & $\gamma^{\prime}_{\mathrm{min}}$  & $1.5$ & $3.0$ & $3.0$ \\
Break Electron Lorentz Factor & $\gamma^{\prime}_{\mathrm{brk}}$ & $7.5\times10^3$ & $7.5\times10^3$ & $7.5\times10^3$  \\
Maximum Electron Lorentz Factor & $\gamma^{\prime}_{\mathrm{max}}$  & $5.0\times10^4$ & $5.0\times10^4$ & $5.0\times10^4$ \\
Low-Energy Electron Spectral Index & $p_1$       & 2.0 & 2.0 & 2.0     \\
High-Energy Electron Spectral Index  & $p_2$       & 4.8 & 4.8 & 4.8	 \\
Magnetic Field [G] & $B^{\prime}$& 0.20  & 0.18 & 0.18    \\
Electron normalization [cm$^{-3}$] & $K^{\prime}$& $3.0\times10^{3}$ & $3.9\times10^{3}$ & $3.7\times10^{3}$ \\
Comoving Radius of the Blob [cm] & $R^{\prime}_{\mathrm{b}}$& 5.8$\times$10$^{16}$ & 6.0$\times$10$^{16}$ & 6.0$\times$10$^{16}$ \\
Distance [cm] & d &5.8$\times$10$^{17}$ & 6.0$\times$10$^{17}$ & 6.0$\times$10$^{17}$ \\
Doppler Factor & $\delta_{\rm D}$      & 23.0 & 20.9 & 17.6    \\
Bulk Lorentz Factor                   & $\Gamma_b$  & 15 & 12 & 10	  \\
\hline
\end{tabular}
\end{table*}

\section{Conclusions}
\label{sec:conclusions}
MAGIC telescopes detected a VHE gamma-ray signal from TON\,0599 on 2017 December 15, following a hardening of the HE gamma-ray spectrum observed with {\textit Fermi}-LAT. With a redshift of 0.7247, TON\,0599 is currently the sixth farthest VHE gamma-ray source, filling a gap in the redshift distribution of the VHE gamma ray emitters (see, e.g., \citealt{Acciari2019}).
The MAGIC telescopes observed TON\,0599 until 2017 December 29, witnessing a gradual fade out of the flare. Simultaneous observations with \textit{Fermi}-LAT showed a smooth transition from HE to VHE gamma-ray spectra, best fitted with an EPWL. The cut-off occurs at energies above 20\,GeV, indicating that the gamma-ray emission region is beyond the BLR, while the dominant reason for the cut-off is most probably the effect of the Klein-Nishina regime.
MWL behaviour was investigated both for the MAGIC observation window (15 days) and a longer time period of 130 days, centred on the MAGIC observations. 
Inspection of the correlation between simultaneous flux measurements in different wavelengths detected strong positive correlation in different bands. 
The variability of the radio flux, unlike other bands, is almost non-existent, both for short and long time periods. 
The correlation study naturally leads to the assumption that a single-zone model in general describes well the long-term data collected. However, the lack of correlation between radio and other bands and the presence of isolated flares (e.g. in the $R$ band) indicate the existence of additional processes not included in the model. 

We modelled the broadband emission independently for each of the four periods of the MAGIC observations. The first two periods are described with the same sets of model parameters, while the last two required a different bulk Lorentz factor to fit the measurements. 
A simple one-zone leptonic model is used, where the low-energy peak is modelled as synchrotron emission from relativistic electrons, whereas the high-energy peak is predominantly produced by EC scattering of photons from the dusty torus. 
The emission region is placed outside of the BLR at a distance consistent with the values previously reported in the literature. However, some parameter values, such as the emission zone radius and magnetic field, are substantially different from those used in studies that did not include measurements in the VHE gamma-ray band.

\section*{Acknowledgements}
We would like to thank the Instituto de Astrof\'{\i}sica de Canarias for the excellent working conditions at the Observatorio del Roque de los Muchachos in La Palma. The financial support of the German BMBF, MPG and HGF; the Italian INFN and INAF; the Swiss National Fund SNF; the grants PID2019-104114RB-C31, PID2019-104114RB-C32, PID2019-104114RB-C33, PID2019-105510GB-C31, PID2019-107847RB-C41, PID2019-107847RB-C42, PID2019-107847RB-C44, PID2019-107988GB-C22, PID2022-136828NB-C41, PID2022-137810NB-C22, PID2022-138172NB-C41, PID2022-138172NB-C42, PID2022-138172NB-C43, PID2022-139117NB-C41, PID2022-139117NB-C42, PID2022-139117NB-C43, PID2022-139117NB-C44 funded by the Spanish MCIN/AEI/ 10.13039/501100011033 and ``ERDF A way of making Europe''; the Indian Department of Atomic Energy; the Japanese ICRR, the University of Tokyo, JSPS, and MEXT; the Bulgarian Ministry of Education and Science, National RI Roadmap Project DO1-400/18.12.2020 and the Academy of Finland grant nr. 320045 is gratefully acknowledged. This work was also been supported by Centros de Excelencia ``Severo Ochoa'' y Unidades ``Mar\'{\i}a de Maeztu'' program of the Spanish MCIN/AEI/ 10.13039/501100011033 (CEX2019-000920-S, CEX2019-000918-M, CEX2021-001131-S) and by the CERCA institution and grants 2021SGR00426 and 2021SGR00773 of the Generalitat de Catalunya; by the Croatian Science Foundation (HrZZ) Project IP-2022-10-4595 and the University of Rijeka Project uniri-prirod-18-48; by the Deutsche Forschungsgemeinschaft (SFB1491) and by the Lamarr-Institute for Machine Learning and Artificial Intelligence; by the Polish Ministry Of Education and Science grant No. 2021/WK/08; and by the Brazilian MCTIC, CNPq and FAPERJ.
T.T. acknowledges the financial support from the University of Rijeka Project uniri-iskusni-prirod-23-24.

Partly based on data taken and assembled by the WEBT collaboration and stored in the WEBT archive at the Osservatorio Astrofisico di Torino-INAF (https://www.oato.inaf.it/blazars/webt/). Partly based on observations made with the Nordic Optical Telescope, owned in collaboration by the University of Turku and Aarhus University, and operated jointly by Aarhus University, the University of Turku and the University of Oslo, representing Denmark, Finland and Norway, the University of Iceland and Stockholm University at the Observatorio del Roque de los Muchachos, La Palma, Spain, of the Instituto de Astrof\'{\i}sica de Canarias.
  Partly based on observations made with the IAC80 operated on the island of Tenerife by the Instituto de Astrof\'{\i}sica de Canarias in the Spanish Observatorio del Teide. Many thanks are due to the IAC support astronomers and telescope operators for supporting the observations at the IAC80 telescope. Partly based on data obtained with the STELLA robotic telescopes in Tenerife, an AIP facility jointly operated by AIP and IAC. The Liverpool Telescope is operated on the island of La Palma by Liverpool John Moores University in the Spanish Observatorio del Roque de los Muchachos of the Instituto de Astrof\'{\i}sica de Canarias with financial support from the UK Science and Technology Facilities Council.
  The Submillimeter Array is a joint project between the Smithsonian Astrophysical Observatory and the Academia Sinica Institute of Astronomy and Astrophysics and is funded by the Smithsonian Institution and the Academia Sinica. We recognize that Maunakea is a culturally important site for the indigenous Hawaiian people; we are privileged to study the cosmos from its summit. The Medicina and Noto radio telescopes are funded by the Ministry of University and Research (MUR) and are operated as National Facility by the National Institute for Astrophysics (INAF).
  The POLAMI observations were carried out at the IRAM 30m Telescope. IRAM is supported by INSU/CNRS (France), MPG (Germany) and IGN (Spain).

We acknowledge financial contribution from the agreement ASI-INAF n.\ 2017-14-H.0 and from the contract PRIN-SKA-CTA-INAF 2016.
This research was partially supported by the Bulgarian National Science Fund of the Ministry of Education and Science under grant KP-06-H68/4 (2022). G.D., M.D.J. and O.V. acknowledge support by the Astronomical Station Vidojevica and the Ministry of Science, Technological Development and Innovation of the Republic of Serbia (MSTDIRS) through contract no. 451-03-66/2024-03/200002 made with Astronomical Observatory (Belgrade), by the EC through project BELISSIMA (call FP7-REGPOT-2010-5, No. 256772), the observing and financial grant support from the Institute of Astronomy and Rozhen NAO BAS through the bilateral SANU-BAN joint research project "GAIA astrometry and fast
variable astronomical objects", and support by the SANU project F-187.
Also, this research was supported by the Science Fund of the Republic of
Serbia, grant no. 6775, Urban Observatory of Belgrade - UrbObsBel.
We acknowledge support by Bulgarian National Science Fund under grant DN18-10/2017 and Bulgarian National Roadmap for Research Infrastructure Project
D01-326/04.12.2023 of the  Ministry of Education and Science of the Republic of
Bulgaria. We thank iTelescope for the use of their telescopes that made it possible for us to carry out optical observations.
The research at Boston University was supported in part by the National Science Foundation grant AST-2108622,  and NASA Fermi Guest Investigator grant  80NSSC23K1507.
This study was based in part on observations conducted using the 1.8m Perkins Telescope Observatory (PTO) in Arizona, which is owned and operated by Boston University.
This publication makes use of data obtained at Mets\"ahovi Radio Observatory, operated by Aalto University in Finland. 
Data from the Steward Observatory spectropolarimetric monitoring project were used. This program is supported by Fermi Guest Investigator grants NNX08AW56G, NNX09AU10G, NNX12AO93G, and NNX15AU81G.
Inna Reva acknowledges financial support  by the Science Committee of the Ministry of Education and Science of the Republic of Kazakhstan under Project No. BR24992759 - ``Development of the concept for the first Kazakhstan orbital cis-lunar telescope - Phase I''. 
The Abastumani team acknowledges financial support by the Shota Rustaveli National Science Foundation of Georgia under contract FR-24-515.

\section*{Affiliations}
{$^{1}$ {Japanese MAGIC Group: Department of Physics, Tokai University, Hiratsuka, 259-1292 Kanagawa, Japan} \\
$^{2}$ {Japanese MAGIC Group: Institute for Cosmic Ray Research (ICRR), The University of Tokyo, Kashiwa, 277-8582 Chiba, Japan} \\
$^{3}$ {ETH Z\"urich, CH-8093 Z\"urich, Switzerland} \\
$^{4}$ {Universit\`a di Siena and INFN Pisa, I-53100 Siena, Italy} \\
$^{5}$ {Institut de F\'isica d'Altes Energies (IFAE), The Barcelona Institute of Science and Technology (BIST), E-08193 Bellaterra (Barcelona), Spain} \\
$^{6}$ {Universitat de Barcelona, ICCUB, IEEC-UB, E-08028 Barcelona, Spain} \\
$^{7}$ {Instituto de Astrof\'isica de Andaluc\'ia-CSIC, Glorieta de la Astronom\'ia s/n, 18008, Granada, Spain} \\
$^{8}$ {National Institute for Astrophysics (INAF), I-00136 Rome, Italy} \\
$^{9}$ {Universit\`a di Udine and INFN Trieste, I-33100 Udine, Italy} \\
$^{10}$ {Max-Planck-Institut f\"ur Physik, D-85748 Garching, Germany} \\
$^{11}$ {Universit\`a di Padova and INFN, I-35131 Padova, Italy} \\
$^{12}$ {Croatian MAGIC Group: University of Zagreb, Faculty of Electrical Engineering and Computing (FER), 10000 Zagreb, Croatia} \\
$^{13}$ {Centro Brasileiro de Pesquisas F\'isicas (CBPF), 22290-180 URCA, Rio de Janeiro (RJ), Brazil} \\
$^{14}$ {IPARCOS Institute and EMFTEL Department, Universidad Complutense de Madrid, E-28040 Madrid, Spain} \\
$^{15}$ {Instituto de Astrof\'isica de Canarias and Dpto. de  Astrof\'isica, Universidad de La Laguna, E-38200, La Laguna, Tenerife, Spain} \\
$^{16}$ {University of Lodz, Faculty of Physics and Applied Informatics, Department of Astrophysics, 90-236 Lodz, Poland} \\
$^{17}$ {Centro de Investigaciones Energ\'eticas, Medioambientales y Tecnol\'ogicas, E-28040 Madrid, Spain} \\
$^{18}$ {Departament de F\'isica, and CERES-IEEC, Universitat Aut\`onoma de Barcelona, E-08193 Bellaterra, Spain} \\
$^{19}$ {Universit\`a di Pisa and INFN Pisa, I-56126 Pisa, Italy} \\
$^{20}$ {INFN MAGIC Group: INFN Sezione di Bari and Dipartimento Interateneo di Fisica dell'Universit\`a e del Politecnico di Bari, I-70125 Bari, Italy} \\
$^{21}$ {Department for Physics and Technology, University of Bergen, Norway} \\
$^{22}$ {INFN MAGIC Group: INFN Sezione di Torino and Universit\`a degli Studi di Torino, I-10125 Torino, Italy} \\
$^{23}$ {Croatian MAGIC Group: University of Rijeka, Faculty of Physics, 51000 Rijeka, Croatia} \\
$^{24}$ {Universit\"at W\"urzburg, D-97074 W\"urzburg, Germany} \\
$^{25}$ {Technische Universit\"at Dortmund, D-44221 Dortmund, Germany} \\
$^{26}$ {Japanese MAGIC Group: Physics Program, Graduate School of Advanced Science and Engineering, Hiroshima University, 739-8526 Hiroshima, Japan} \\
$^{27}$ {Deutsches Elektronen-Synchrotron (DESY), D-15738 Zeuthen, Germany} \\
$^{28}$ {Armenian MAGIC Group: ICRANet-Armenia, 0019 Yerevan, Armenia} \\
$^{29}$ {Croatian MAGIC Group: University of Split, Faculty of Electrical Engineering, Mechanical Engineering and Naval Architecture (FESB), 21000 Split, Croatia} \\
$^{30}$ {Croatian MAGIC Group: Josip Juraj Strossmayer University of Osijek, Department of Physics, 31000 Osijek, Croatia} \\
$^{31}$ {Finnish MAGIC Group: Finnish Centre for Astronomy with ESO, Department of Physics and Astronomy, University of Turku, FI-20014 Turku, Finland} \\
$^{32}$ {University of Geneva, Chemin d'Ecogia 16, CH-1290 Versoix, Switzerland} \\
$^{33}$ {Saha Institute of Nuclear Physics, A CI of Homi Bhabha National Institute, Kolkata 700064, West Bengal, India} \\
$^{34}$ {Inst. for Nucl. Research and Nucl. Energy, Bulgarian Academy of Sciences, BG-1784 Sofia, Bulgaria} \\
$^{35}$ {Japanese MAGIC Group: Department of Physics, Yamagata University, Yamagata 990-8560, Japan} \\
$^{36}$ {Finnish MAGIC Group: Space Physics and Astronomy Research Unit, University of Oulu, FI-90014 Oulu, Finland} \\
$^{37}$ {Japanese MAGIC Group: Chiba University, ICEHAP, 263-8522 Chiba, Japan} \\
$^{38}$ {Japanese MAGIC Group: Institute for Space-Earth Environmental Research and Kobayashi-Maskawa Institute for the Origin of Particles and the Universe, Nagoya University, 464-6801 Nagoya, Japan} \\
$^{39}$ {Japanese MAGIC Group: Department of Physics, Kyoto University, 606-8502 Kyoto, Japan} \\
$^{40}$ {INFN MAGIC Group: INFN Roma Tor Vergata, I-00133 Roma, Italy} \\
$^{41}$ {Japanese MAGIC Group: Department of Physics, Konan University, Kobe, Hyogo 658-8501, Japan} \\
$^{42}$ {also at International Center for Relativistic Astrophysics (ICRA), Rome, Italy} \\
$^{43}$ {also at Como Lake centre for AstroPhysics (CLAP), DiSAT, Universit\`a dell?Insubria, via Valleggio 11, 22100 Como, Italy.} \\
$^{44}$ {also at Port d'Informaci\'o Cient\'ifica (PIC), E-08193 Bellaterra (Barcelona), Spain} \\
$^{45}$ {now at Universit\'e Paris Cit\'e, CNRS, Astroparticule et Cosmologie, F-75013 Paris, France} \\
$^{46}$ {also at Dipartimento di Fisica, Universit\`a di Trieste, I-34127 Trieste, Italy} \\
$^{47}$ {Max-Planck-Institut f\"ur Physik, D-85748 Garching, Germany} \\
$^{48}$ {also at INAF Padova} \\
$^{49}$ {Japanese MAGIC Group: Institute for Cosmic Ray Research (ICRR), The University of Tokyo, Kashiwa, 277-8582 Chiba, Japan} \\
$^{50}$ {Crimean Astrophysical Observatory RAS, P/O Nauchny, 298409, Crimea} \\
$^{51}$ {Fundaci\'on Galileo Galilei - INAF (Telescopio Nazionale Galileo), Rambla Jos\'e Ana Fern\'andez Perez 7, E-38712 Bre\~na Baja (La Palma), Canary Islands, Spain} \\
$^{52}$ {Department of Astronomy, Faculty of Physics, Sofia, Bulgaria} \\
$^{53}$ {University ``St Kliment Ohridski'', 5 James Bourchier Blvd., BG-1164 Sofia, Bulgaria} \\
$^{54}$ {INAF-Osservatorio Astrofisico di Torino, via Osservatorio 20, I-10025 Pino Torinese, Italy} \\
$^{55}$ {EPT Observatories, Tijarafe, La Palma, Spain} \\
$^{56}$ {Foundation for Research and Technology - Hellas, IESL \& Institute of Astrophysics, Voutes, 7110, Heraklion, Greece; Department of Physics, University of Crete, 70013, Heraklion, Greece} \\
$^{57}$ {Helmholtz centre Potsdam, GFZ German Research Centre for Geosciences, Germany} \\
$^{58}$ {National Central University, 300 Zhongda Road, Zhongli 32001, Taoyuan, Taiwan} \\
$^{59}$ {Astronomical Observatory, Volgina 7, 11060 Belgrade, Serbia} \\
$^{60}$ {Ulugh Beg Astronomical Institute, Astronomy Street 33, Tashkent 100052, Uzbekistan} \\
$^{61}$ {National University of Uzbekistan, Tashkent 100174, Uzbekistan} \\
$^{62}$ {INAF --- Istituto di Radioastronomia, Via Gobetti 101, 40129 Bologna, Italy} \\
$^{63}$ {Saint Petersburg State University, 7/9 Universitetskaya nab., St. Petersburg, 199034 Russia} \\
$^{64}$ {Center for Astrophysics, Harvard \& Smithsonian, 60 Garden Street, Cambridge, MA 02138 USA} \\
$^{65}$ {Institute for Astrophysical Research, Boston University, 725 Commonwealth Avenue, Boston, MA 02215, USA} \\
$^{66}$ {Department of Physics \& Astronomy, Texas Tech University, Box 41051, Lubbock, TX, 79409-1051, USA} \\
$^{67}$ {Astrophysics Research Institute, Liverpool John Moores University, IC2, Liverpool Science Park, Liverpool, L3 5RF, UK} \\
$^{68}$ {National Center of Space Researches and Technologies, Almaty, Kazakhstan} \\
$^{69}$ {Al-Farabi Kazakh National University, Almaty, Kazakhstan} \\
$^{70}$ {Abastumani Observatory, Mt. Kanobili, 0301 Abastumani, Georgia} \\
$^{71}$ {Max-Planck-Institut f\"ur Radioastronomie, Auf dem H\"ugel 69, 53121 Bonn, Germany} \\
$^{72}$ {Landessternwarte, Zentrum fur Astronomie der Universit\"at Heidelberg, K\"onigstuhl 12, 69117 Heidelberg, Germany} \\
$^{73}$ {Engelhardt Astronomical Observatory, Kazan Federal University, Tatarstan, Russia} \\
$^{74}$ {Aalto University Mets\"ahovi Radio Observatory, Mets\"ahovintie 114, 02540 Kylm\"al\"a, Finland} \\
$^{75}$ {Aalto University Department of Electronics and Nanoengineering, P.O. BOX 15500, FI-00076 AALTO, Finland} \\
$^{76}$ {Pulkovo Observatory, St. Petersburg, 196140, Russia} \\
$^{77}$ {Institute of Astronomy and National Astronomical Observatory, Bulgarian Academy of Sciences, 72 Tsarigradsko shosse Blvd., 1784 Sofia, Bulgaria} \\
$^{78}$ {Institut de Radiostronomie Milim\'etrique, Avenida Divina Pastora, 7, Local 20, E-18012 Granada, Spain} \\
$^{79}$ {Nordic Optical Telescope,  Apartado 474, E-38700 Santa Cruz de La Palma, Santa Cruz de Tenerife, Spain} \\
$^{80}$ {Department of Physics and Astronomy, Aarhus University, Munkegade 120, DK-8000 Aarhus C, Denmark} \\
$^{81}$ {Fesenkov Astrophysical Institute, Observatory street 23, Almaty 050020, Kazakhstan} \\
$^{82}$ {Department of Physics, University of Colorado, Denver, USA} \\
$^{83}$ {Steward Observatory, University of Arizona, 933 N. Cherry Ave., Tucson, AZ 85721 USA} \\
$^{84}$ {Istituto di Astrofisica e Planetologia Spaziali -- INAF, Via Fosso del Cavaliere, 100 -- I-00133 Rome, Italy}
}

\section*{Data Availability}
The data underlying this article will be shared on reasonable request to the corresponding authors.




\bibliographystyle{mnras}
\bibliography{TON0599-MAGIC_MNRAS} 



\appendix

\section*{Author Contributions}
T. Terzi\'{c} project leadership, paper organisation, drafting, and editing; A. Stamerra triggering MWL observations; T. Terzi\'{c} and J. Hirako MAGIC data analysis, L. Pacciani \textit{Fermi}-LAT data analysis and paper drafting, C. Righi \textit{Swift} data analysis. F. Tavecchio and C. Righi emission modelling and paper drafting. C.M. Raiteri and M. Villata for the WEBT campaign management, WEBT data analysis and paper drafting. T. Terzi\'{c} light curve variability and correlation analysis. The rest of the authors have contributed in one or several of the following ways: design, construction, maintenance and operation of the instrument(s) used to acquire the data; preparation and/or evaluation of the observation proposals; data acquisition, processing, calibration and/or reduction; production of analysis tools and/or related Monte Carlo simulations; discussion and approval of the contents of the draft.


\bsp	
\label{lastpage}
\end{document}